# Illuminating the Future: Nanophotonics for Future Green Technologies, Precision Healthcare, and Optical Computing

Osama M. Halawa,[a] Esraa Ahmed,[b] Malk M. Abdelrazek,[c] Yasser M. Nagy,[c] Omar A. M. Abdelraouf, [d, *]

[a] Mechanical Engineering department, Faculty of Engineering at El-Mataria, Helwan University, Cairo, Egypt

[b] Laser Institute for Research and Applications LIRA, Beni-Suef University, Beni-Suef 62511, Egypt

[c] Beni-suef students research unit, Faculty of Medicine, Beni-Suef University, Egypt

[d] Institute of Materials Research and Engineering, Agency for Science, Technology, and Research (A*STAR), 2 Fusionopolis Way, #08-03, Innovis, Singapore 138634, Singapore.

* Corresponding author. Email address: Omar_Abdelrahman@imre.a-star.edu.sg

## Abstract

Nanophotonics, an interdisciplinary field merging nanotechnology and photonics, has enabled transformative advancements across diverse sectors including green energy, biomedicine, and optical computing. This review comprehensively examines recent progress in nanophotonic principles and applications, highlighting key innovations in material design, device engineering, and system integration. In renewable energy, nanophotonic allows light-trapping nanostructures and spectral control in perovskite solar cells, concentrating solar power, and thermophotovoltaics. That have significantly enhanced solar conversion efficiencies, approaching theoretical limits. For biosensing, nanophotonic platforms achieve unprecedented sensitivity in detecting biomolecules, pathogens, and pollutants, enabling real-time diagnostics and environmental monitoring. Medical applications leverage tailored light-matter interactions for precision photothermal therapy, image-guided surgery, and early disease detection. Furthermore, nanophotonics underpins next-generation optical neural networks and neuromorphic computing, offering ultra-fast, energy-efficient alternatives to von Neumann architectures. Despite rapid growth, challenges in scalability, fabrication costs, and material stability persist. Future advancements will rely on novel materials, AI-driven design optimization, and multidisciplinary approaches to enable scalable, low-cost deployment. This review summarizes recent progress and highlights future trends, including novel material systems, multidisciplinary approaches, and enhanced computational capabilities, to pave the way for transformative applications in this rapidly evolving field.

# 1 Introduction

## 1.1 Overview of Nanophotonics and Its Significance

Light, the fundamental carrier of information and energy, has been harnessed by humanity for millennia through traditional optics, employing lenses, mirrors, and prisms crafted from bulk materials. While foundational to scientific progress, these macroscopic components face inherent limitations as technology advances towards miniaturization, enhanced efficiency, and novel functionalities. Diffraction fundamentally constrains resolution and device size, the physical scale hinders integration into compact systems like semiconductor chips. Diffraction weak light-matter interaction in bulk media necessitates high power or long path lengths for significant nonlinear effects or efficient sensing. Nanophotonics emerges as the transformative solution to these challenges. This vibrant, interdisciplinary field, operating at the intersection of optics, photonics, materials science, nanotechnology, and quantum physics, focuses on manipulating light using structures and devices with critical features on the scale of, or smaller than, the wavelength of light itself, typically tens to hundreds of nanometers.[1-3]

By engineering materials and geometries at this subwavelength scale, nanophotonics circumvents the diffraction limit and enable unprecedented control over light propagation, localization, emission, and absorption.[4] This mastery unlocks phenomena impossible or highly inefficient in bulk optics: confining light to volumes far smaller than its wavelength, sculpting optical wavefronts with ultra-thin components, and dramatically enhancing light-matter interactions. These capabilities represent not merely incremental improvements but a paradigm shift, paving the way for revolutionary devices and applications across diverse sectors, including sustainable energy, life-saving healthcare, ultra-fast computing, and quantum technologies.[5-7]

The profound advantages of nanophotonics stem directly from its ability to transcend classical optical constraints and exploit unique nanoscale physics. Critically, it breaks the diffraction limit; structures like plasmonic nanoantennas or high-index dielectric resonators concentrate optical energy into "hot spots" with dimensions significantly smaller than half the wavelength. That enables super-resolution imaging, ultra-compact components, and molecular-level probing. Furthermore, it facilitates ultra-compact integration. By replacing bulky lenses and mirrors with sub-micron planar devices like metasurfaces allows complex optical systems to be miniaturized and integrated directly onto semiconductor chips, enabling photonic integrated circuits (PICs) and lab-on-a-chip platforms. Perhaps most significantly, nanophotonics dramatically enhances light-matter interactions. Through confining light to extremely small volumes increases local electromagnetic field intensity and photon density of states and profoundly boosting absorption for photodetectors and solar cells. Also, by enhancing the emission rate of quantum emitters for brighter sources,[8-11] lowering power thresholds for nonlinear optical effects like harmonic generation,[12-17] and amplifying sensitivity to minute environmental changes for advanced sensing.[18-22] Additionally, it enables the tailored engineering of optical properties, creating metamaterials with exotic characteristics like negative refraction and metasurfaces with spatially varying responses for complex wavefront shaping, all while consuming orders of magnitude less material than bulk optics, offering benefits in cost, sustainability, and weight.

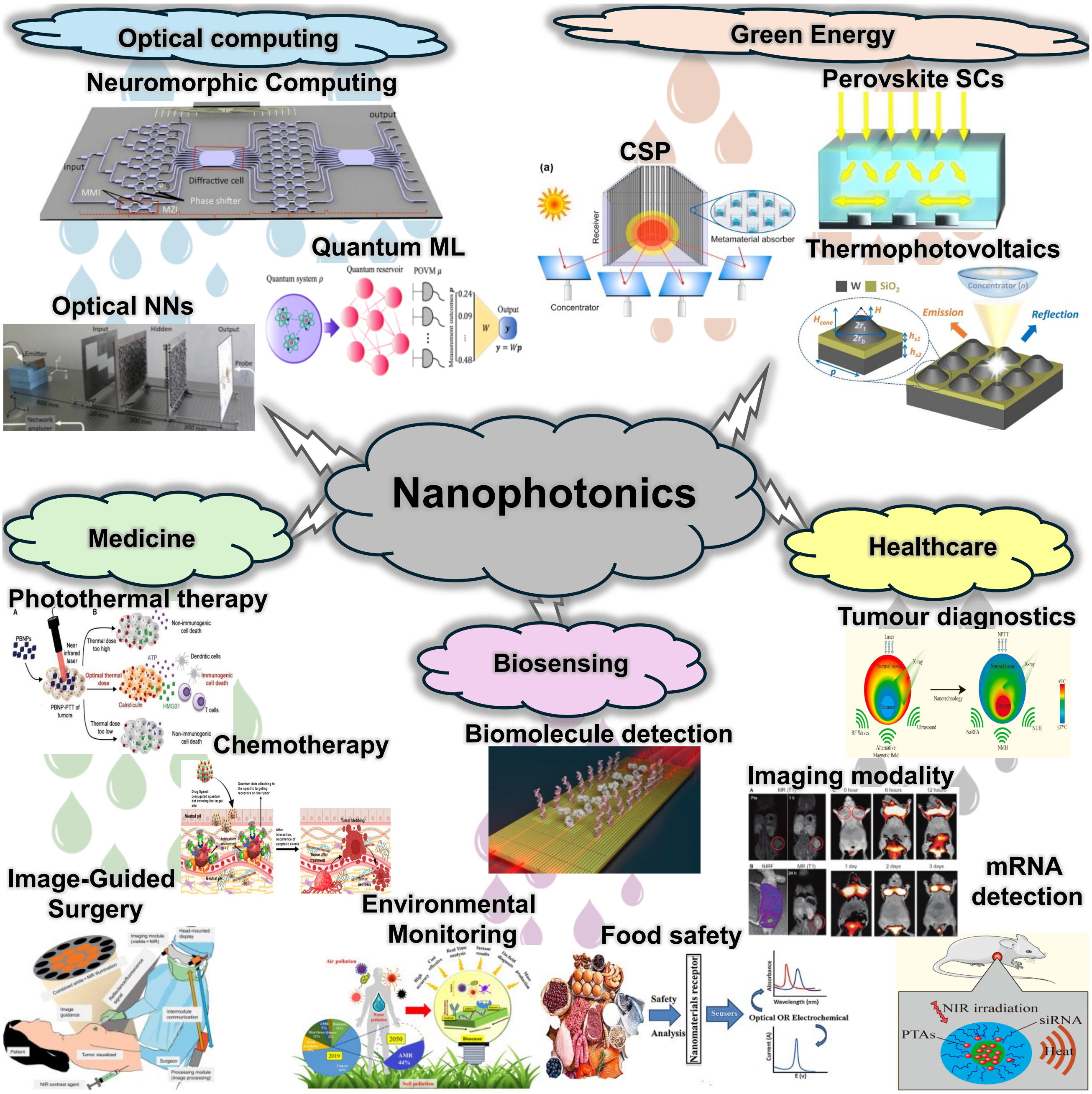

**Figure 1:** Nanophotonics applications clouds. Reprinted with permission from.[23-35]

## 1.2 Fundamental Physics of Nanophotonics

The foundation of nanophotonics lies in meticulously designed nanostructures governed by fundamental physical principles. A revolutionary concept is the metasurface, an artificial sheet material, nanometers to a few hundred nanometers thick, composed of dense arrays of subwavelength nanostructures (meta-atoms). By precisely designing the geometry, material, and arrangement of these meta-atoms, local control over the amplitude, phase, and polarization of scattered light is achieved. A cornerstone achievement is the phase gradient metasurface. By imposing a linear spatial gradient in the phase shift imparted by adjacent meta-atoms, these devices anomalously refract or reflect incident light according to the Generalized Laws of Reflection and Refraction. Unlike Snell's law governing homogeneous bulk interfaces, the generalized laws predict deflection angles dependent on both the refractive index difference and the deliberately introduced phase gradient.[36-38]

$$\sin(\theta_r) - \sin(\theta_i) = \frac{\lambda_o}{2\pi n_i}\frac{d\phi}{dx} \tag{1}$$

$$\sin(\theta_t)\, n_t - \sin(\theta_i)\, n_I = \frac{\lambda_o}{2\pi}\frac{d\phi}{dx} \tag{2}$$

Where $n_i$, $n_t$ are incident and transmitted refractive indices, $\theta_i$, $\theta_t$ are incident and transmitted angles, $\lambda_0$ is the free-space wavelength, and $\frac{d\phi}{dx}$ is the phase gradient. This principle enables ultra-thin, flat lenses (metalenses), beam steerers, holograms, and polarizers, replacing bulky optics with planar, CMOS-compatible components capable of sophisticated wavefront manipulation. Achieving deep subwavelength light confinement and intense field enhancement relies heavily on resonant phenomena. Mie scattering theory provides the fundamental framework for understanding electromagnetic wave interactions with subwavelength dielectric objects (e.g., silicon, GaAs nanoparticles). High-index dielectrics support Mie resonances including magnetic dipole (MD, from circular displacement currents), electric dipole (ED, from oscillating polarization), and higher-order multipoles. The governing equations for diploe and quadrupole optical modes are:[39-41]

$$C_{ED} = \frac{k_0^4}{6\,\pi\epsilon_0^2 E_0^2}\left|p_{car} + \frac{ik_0}{c}\left(t + \frac{k_0^2}{10}\overline{R_t^2}\right)\right|^2 \tag{3}$$

$$C_{EQ} = \frac{k_0^6}{80\,\pi\epsilon_0^2 E_0^2}\left|\overline{\overline{Q_e}} + \frac{ik_0}{c}\overline{\overline{Q_t}}\right|^2 \tag{4}$$

$$C_{MD} = \frac{\eta_0^2 k_0^4}{6\,\pi E_0^2}\left|m_{car} - k_0^2\overline{R_m^2}\right|^2 \tag{5}$$

$$C_{MQ} = \frac{\eta_0^2 k_0^6}{80\,\pi E_0^2}\left|\overline{\overline{Q_m}}\right|^2 \tag{6}$$

where $C_{ED}$, $C_{MD}$, $C_{EQ}$, and $C_{MQ}$ denote the scattering cross-section of an dipole modes for electric and magnetic fields, and quadrupolar modes for electric and magnetic fields, respectively. Optical modes arising from displacement currents within the material and offering low optical losses compared to plasmonic metals. Multi-optical mode metasurfaces represent a sophisticated evolution, where nanostructures are engineered to simultaneously support multiple, spectrally distinct Mie resonances. This multi-modal capability is pivotal for nonlinear optics, where resonances at both fundamental pump and generated harmonic wavelengths boost interaction strength; specific modal overlaps can enable or enhance processes and facilitate unique phase-matching, while extreme field confinement amplifies nonlinear coefficients. It is equally crucial for quantum optics, where resonant structures enhance the spontaneous emission rate of quantum emitters (Purcell effect) via tailored modal profiles. Designing multi-mode metasurface will enable control emission directionality, polarization, and efficiency, and enable strong light-matter coupling for quantum information processing and nonlinear quantum optics at the single-photon level.[42-44]

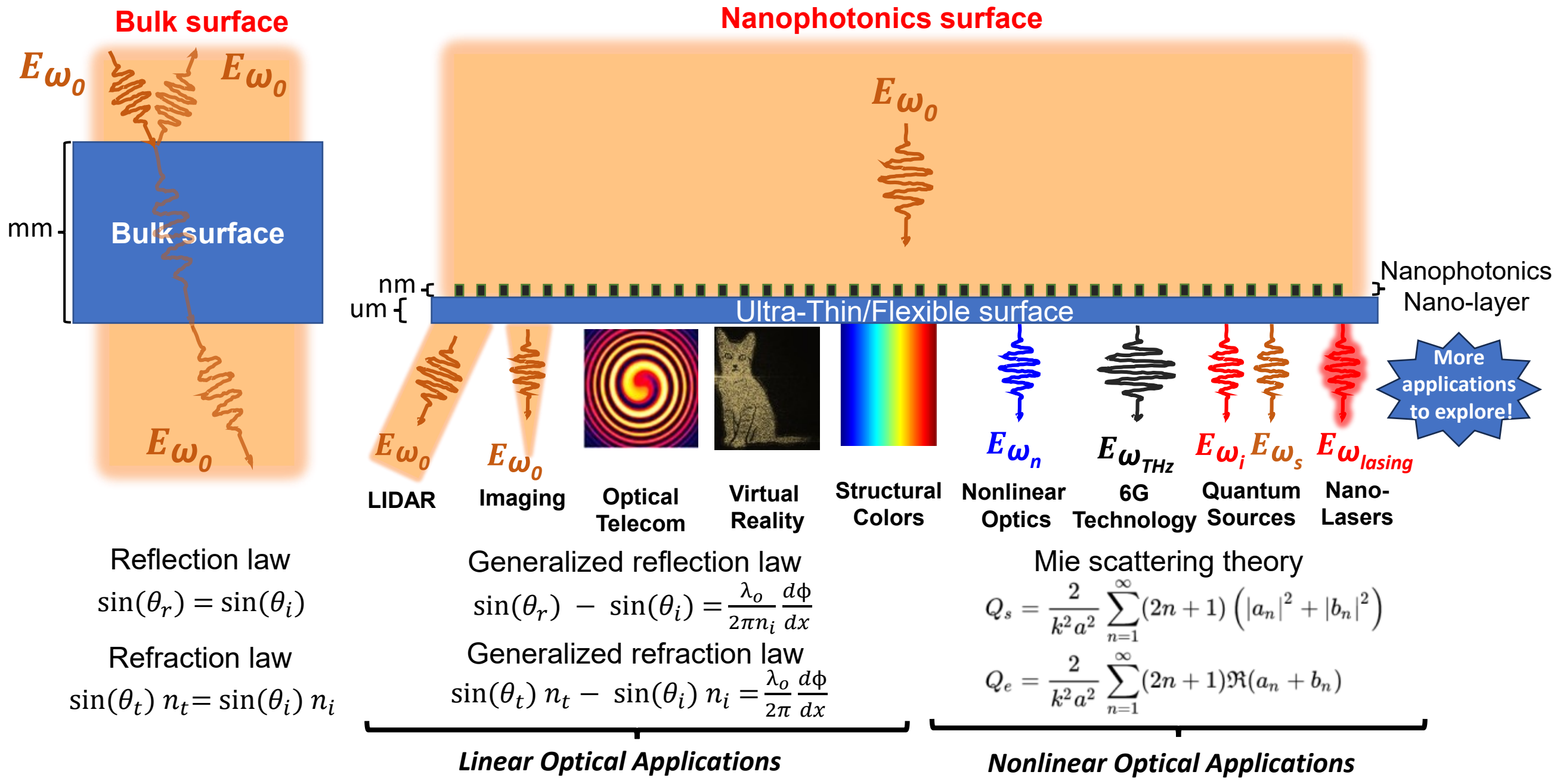

**Figure 2:** Nanophotonics vs bulk photonic devices.

The unique capabilities unlocked by nanophotonic design principles are actively driving transformative solutions across critical global challenges and enabling novel functionalities. In the realm of green energy, nanophotonics significantly enhances efficiency. For photovoltaics, nanostructured surfaces minimize reflection losses, while light-trapping schemes using nanostructures increase the effective optical path length within absorber layers, boosting absorption for thin-film technologies; spectral splitting better matches sunlight to multi-junction cell bandgaps, pushing efficiencies closer to theoretical limits with reduced material costs. In photocatalysis, plasmonic nanoparticles act as nanoantennas concentrating light to generate hot electrons for driving reactions like water splitting, while dielectric nanostructures enhance absorption within semiconductors and improve charge separation, enabling efficient systems for sustainable fuel production and environmental remediation.[45]

Biosensing and environmental monitoring benefit tremendously from accurate and noninvasive inspection. Label-free biosensing leverages the exquisite refractive index sensitivity of nanophotonic resonators (plasmonic nanostructures, photonic crystals, dielectric metasurfaces); binding of target biomolecules to functionalized surfaces induces detectable spectral shifts, enabling real-time, highly sensitive, multiplexed detection for disease diagnosis and point-of-care testing. Similarly, functionalized nanophotonic sensors provide real-time, highly sensitive, and selective monitoring of pollutants in air and water, facilitated by miniaturized, rugged sensor nodes for widespread environmental surveillance. Healthcare sees profound impacts in diagnosis and safe, targeted treatment. Advanced imaging leverages nanophotonics for super-resolution microscopy breaking the diffraction barrier and enhancing contrast in techniques like OCT, while nanophotonic biosensors offer rapid diagnostics. Crucially, safe cancer optothermal therapy utilizes plasmonic nanoparticles (e.g., gold nanorods) engineered to absorb near-infrared light within the "biological window"; injected into tumors and illuminated, they efficiently convert light into localized heat, selectively ablating cancerous cells with minimal damage to healthy tissue, offering a minimally invasive alternative.[32]

Recently, nanophotonics is powering the next generation of computing. For high-speed interconnects, nanoscale waveguides, modulators, and detectors on silicon chips enable ultra-fast, low-power optical communication between processor cores and memory, alleviating the "von Neumann bottleneck." In neuromorphic computing, nanophotonic components like metasurfaces and programmable photonic circuits perform complex linear operations (matrix multiplications) inherent in artificial neural networks at the speed of light with minimal energy; nonlinear nanophotonic elements implement activation functions, promising revolutionary gains in speed and efficiency for AI and complex data analysis.[46]

Nanophotonics stands as a defining technology of the modern era, fundamentally reshaping our mastery over light and its interaction with matter. By overcoming the diffraction barrier through nanoscale engineering, it unlocks unparalleled capabilities. The extreme light confinement, precise wavefront control with ultra-thin devices, and massively amplified light-matter interactions. The foundational physics of generalized refraction enabled by phase-gradient metasurfaces and the intricate multi-resonant phenomena governed by Mie theory in dielectric nanostructures, provides a powerful design framework. The following sections of this review will delve deeper into boosting green energy efficiency, enabling ultra-sensitive biosensing and environmental monitoring, pioneering safe cancer therapies and efficient healthcare, and laying the groundwork for next-generation optical computing.

# 2 Nanophotonics for Energy Production and Conversion

The quest for efficient and cost-effective solar energy conversion has driven remarkable innovations in nanophotonics over the past decade. Nanophotonics, the study of light behavior on the nanometer scale, has enabled breakthrough technologies that manipulate light-matter interactions at the nanoscale to enhance solar energy harvesting.[47-54] This section of the review focuses on three transformative technologies that exemplify the power of nanophotonics in solar energy: perovskite solar cells, concentrating solar power systems, and solar thermophotovoltaics. These technologies represent different approaches to solar energy conversion, each leveraging nanophotonic principles to overcome traditional limitations and achieve unprecedented performance levels. From the rapid efficiency improvements in perovskite solar cells to the enhanced light concentration capabilities in solar power systems, nanophotonics continues to revolutionize how we capture and convert solar energy.

## 2.1 Perovskite Solar Cells

### 2.1.1 Overview of Perovskite Solar Cell Technology

Perovskite solar cells (PSCs) represent one of the most promising and rapidly advancing photovoltaic technologies of the 21st century. A perovskite solar cell is fundamentally a type of solar cell that incorporates a perovskite-structured compound as its light harvesting active layer.[55] The term "perovskite" originates from the mineral perovskite ($CaTiO_3$), discovered in the Ural Mountains of Russia by Gustav Rose in 1839 and named after Russian mineralogist Lev Perovskite.[56]

The defining characteristic of perovskite solar cells lies in their crystal structure, which follows the general formula $AMX_3$, where A is an organic ammonium cation (such as methylammonium or formamidinium), M is a metal cation (typically Pb or Sn), and X is a halide anion.[57] The size of cation A is critical for the formation of a close-packed perovskite structure, as it must fit into the space composed of four adjacent corner-sharing $MX_6$ octahedra. This compositional flexibility enables tuning of material properties through chemical modification of different sites, particularly the A and X positions, which can significantly influence bandgap, phase stability, and overall photovoltaic performance. In the context of solar cells, the most commonly studied perovskite absorber is methylammonium lead trihalide ($CH_3NH_3PbX_3$), where X represents a halogen ion such as iodide, bromide, or chloride.[58] This structure creates a three-dimensional network of corner-sharing octahedra, with the larger A cation occupying the cuboctahedral sites formed by the octahedral framework.

The perovskite crystal structure can be visualized as a cubic unit cell where the B cation (typically lead in photovoltaic applications) occupies the corners, the X anions (halides) are positioned at the midpoints of the cube edges, and the A cation (organic molecules like methylammonium or formamidinium) resides in the center of the cube.[59] This arrangement creates a highly ordered crystalline structure that exhibits remarkable optoelectronic properties essential for photovoltaic applications.

The integration of nanophotonics with perovskite solar cells has emerged as a transformative approach for enhancing device performance through sophisticated light management strategies.[60-62] Nanophotonic techniques enable precise control over light-matter interactions at the nanoscale, addressing key challenges in perovskite devices including light absorption optimization, charge carrier transport enhancement, and optical loss minimization.[63] These approaches encompass nanopatterning, nanotexturing, and nanostructuring methodologies that can significantly improve photon harvesting efficiency while maintaining the inherent advantages of perovskite materials.

The implementation of nanophotonic designs has demonstrated remarkable success in improving perovskite solar cell efficiencies, with recent advances showing progression from 3.8% to over 25.5% for single-junction devices and exceeding 30% for tandem configurations. Furthermore, nanophotonic structures offer unique advantages for perovskite devices by enabling better integration compatibility compared to conventional silicon-based approaches, while simultaneously providing pathways for enhanced stability and reduced manufacturing complexity.

### 2.1.2 Recent Developments in Perovskite Solar Cells

The field of perovskite solar cells has witnessed remarkable progress in recent years, with significant advances in efficiency, stability, device architectures, and novel material engineering approaches. This section examines the latest developments based on cutting-edge research that has pushed the boundaries of perovskite photovoltaic technology toward commercial viability.

#### Advanced Optical Characterization and Performance Enhancement

Recent research has emphasized the critical importance of precise optical characterization in optimizing perovskite solar cell performance. Widianto et al. conducted a comprehensive investigation into the application of spectroscopic ellipsometry (SE) for advancing perovskite solar cell technology.[64] Their work demonstrates that SE provides essential insights into optical constants such as the dielectric function, refractive index, and absorption coefficient of perovskite thin films. The research highlights that understanding these optical characteristics is fundamental to designing improved device structures and enhancing overall performance.

The study reveals that spectroscopic ellipsometry enables systematic evaluation of optoelectronic properties in complex, multilayered perovskite systems. This technique has proven particularly valuable for analyzing temperature-dependent behavior, thermal degradation mechanisms, and chemical composition effects that critically influence both performance and stability of solar cell devices. The ability to precisely characterize these optical properties represents a significant advancement in the field, as it provides researchers with the tools necessary to optimize light collection efficiency, which is essential for enhancing power conversion efficiency in perovskite solar cells.

Furthermore, the research demonstrates that SE-based techniques serve as effective tools for thoroughly evaluating perovskite thin films, focusing on critical aspects such as optical model generation, data analysis methodologies, and the extraction of key optical constants. This comprehensive approach to optical characterization has opened new pathways for accelerating the commercialization of perovskite solar cells by providing deeper insights into the fundamental optical processes that govern device performance.[64]

**Breakthrough Achievements in Tandem Solar Cell Technology**

The development of tandem solar cell architectures represents one of the most promising approaches for surpassing the efficiency limitations of single-junction devices. Qian et al. reported groundbreaking achievements in four-terminal tandem devices through innovative light utilization optimization strategies.[65] Their research focused on semi-transparent perovskite solar modules (ST-PSMs) and introduced a novel approach by incorporating tin oxide nanoparticles ($SnO_2$ NPs) into the perovskite solution.

This innovative strategy led to the construction of p-n homojunctions within the upper layer of large-area films, resulting in significant performance improvements. The research demonstrates that this approach enhances the built-in electric field through pn homojunctions while simultaneously improving the circulation of visible light within the perovskite film via nanoparticle scattering effects. This dual action mechanism has proven highly effective in improving both charge transport efficiency and light management capabilities.

The experimental results achieved by this research team are particularly noteworthy. Their 56.9 $cm^2$ semi-transparent perovskite solar modules achieved a certified power conversion efficiency of 17.2%, representing a significant advancement in large-area device performance. When these modules were mechanically stacked with silicon heterojunction (SHJ) solar cells to form four-terminal tandem devices, the combined system achieved an impressive power conversion efficiency of 27.2%. This achievement demonstrates the substantial potential of tandem architectures for overcoming the theoretical efficiency limits of single-junction devices.

The significance of this work extends beyond the immediate efficiency gains, as it establishes a new paradigm for enhancing the performance of perovskite tandem solar devices. The research provides a promising avenue for future solar energy applications by demonstrating that carefully engineered nanostructures can simultaneously address multiple performance-limiting factors in perovskite devices.

**Revolutionary Advances in Hole Transport Materials**

The development of advanced hole transport materials has emerged as a critical factor in achieving high-efficiency perovskite solar cells. Recent research by the Nanophotonics Team at Guangxi University has yielded a series of groundbreaking achievements in this area. Their work represents a systematic approach to material innovation that has resulted in some of the highest reported efficiencies for perovskite solar cells.

One of the most significant contributions from this research group is the design and synthesis of a novel hole transport material named V3PACz with in-situ crosslinking capability. This breakthrough was achieved by combining an in-situ self-polymerization strategy with carbazole phosphonic acid self-assembled molecules.[66] The innovative approach enabled the fabrication of high-quality Poly-V3PACz hole transport layers, which achieved an outstanding power conversion efficiency of 25.21% along with excellent stability characteristics.

The research team also made significant progress in developing organic dopant small molecules as active layer additives. They designed and synthesized two PDI-based ntype A-D-A structured organic dopant small molecules, PBDT and PTBDT, which significantly improved perovskite film quality, internal charge extraction capability, passivation effects, and overall device stability.[67] The implementation of these compounds in devices with an active area of 0.08 $cm^2$ resulted in a remarkable power conversion efficiency of 25.94% with an open-circuit voltage of 1.18 V and a fill factor of 86.37%. These values represent some of the highest reported open-circuit voltage and power conversion efficiency values for single-junction inverted perovskite solar cells to date.

Perhaps most innovatively, the research team pioneered the use of deuteration in material design strategy by developing a novel self-assembling hole transport material, 4PACzd8. Perovskite solar cells employing 4PACzd8 as the hole transport layer achieved an impressive power conversion efficiency of 24.87% along with exceptional ultraviolet light stability.[68] This study demonstrates the significant value of deuteration strategy in the design of small-molecule hole transport materials, opening new avenues for material engineering in perovskite photovoltaics.

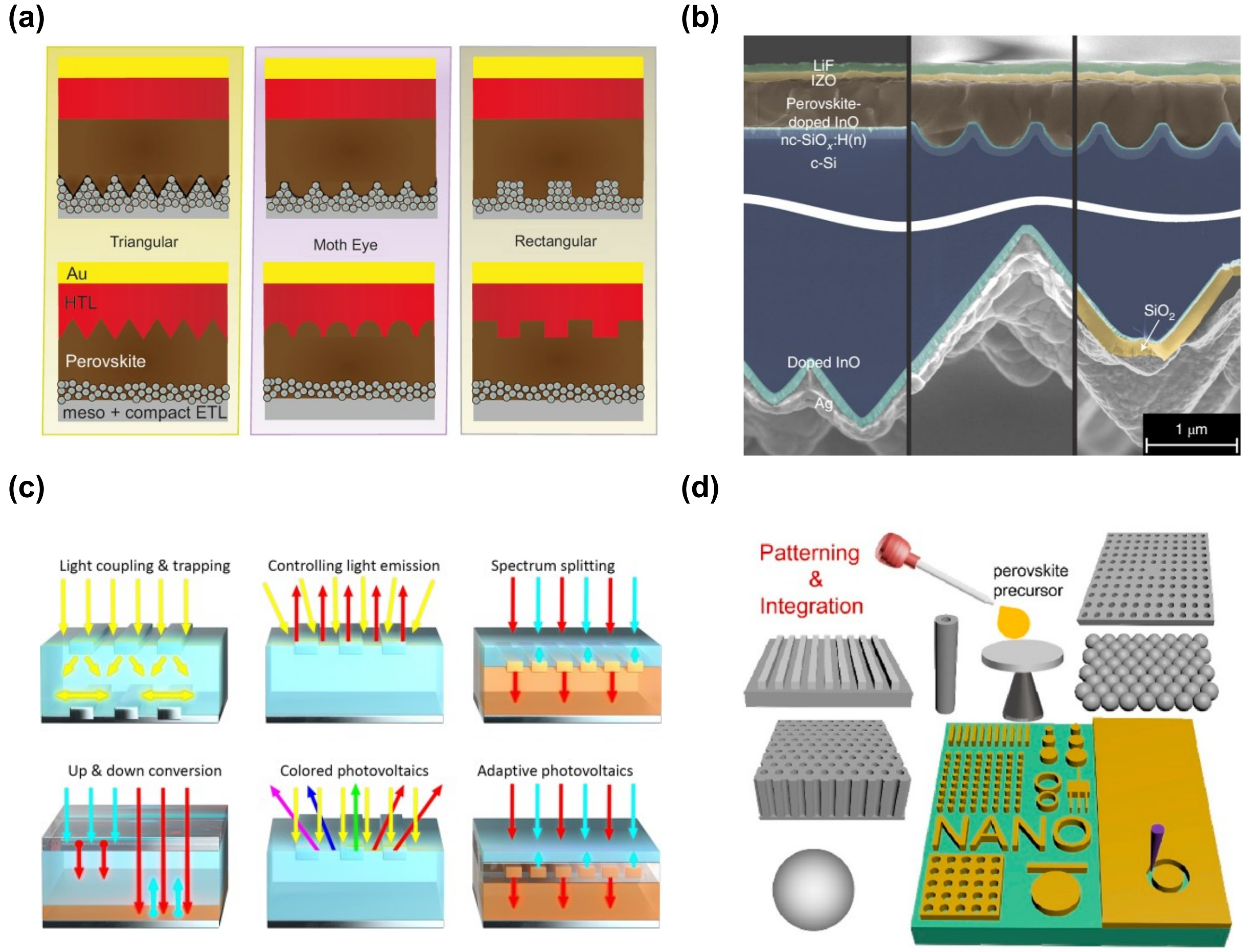

**Figure 3.** Some of the recent advancements in nanophotonics for improving perovskite solar cell performance. (a) nanostructured designs with triangular interfaces, moth-eye architectures, and rectangular interfaces.[63] (b) SEM cross-sections of planar, nanotextured, and nanotextured + RDBL PSTSCs showing front and rear sides, with c-Si indicating crystalline silicon.[69] (c) light-management structures for enhanced photovoltaic efficiency.[23] (d) patterning and integration techniques for micro- and nanostructured perovskites.[70]

## Nano-optical Engineering for Enhanced Performance

The integration of nano-optical designs has emerged as a powerful strategy for improving the efficiency of perovskite-silicon tandem solar cells.[71, 72] Tockhorn et al. presented a comprehensive study on perovskite-silicon tandem solar cells with periodic nanotextures that offer significant advantages without compromising the material quality of solution-processed perovskite layers. Their research addresses one of the fundamental challenges in tandem device development: optimizing optical performance while maintaining high-quality film growth.

The study demonstrates that periodic nanotextures provide substantial benefits compared to conventional planar tandem devices. The nanotextured devices showed a significant reduction in reflection losses while exhibiting reduced sensitivity to deviations from optimum layer thicknesses. This characteristic is particularly important for manufacturing scalability, as it provides greater tolerance for process variations that are inevitable in large-scale production.

One of the most remarkable achievements of this research is the dramatic improvement in fabrication yield. The implementation of nanotextures enabled a greatly increased fabrication yield from 50% to 95%, representing a nearly two-fold improvement in manufacturing success rate. This enhancement is crucial for the commercial viability of perovskite tandem technology, as it directly impacts production costs and scalability.

The research also revealed that nanotextures contribute to improved optoelectronic properties of the perovskite top cell, resulting in a 15-mV improvement in open-circuit voltage. Additionally, the team developed an optically advanced rear reflector with a dielectric buffer layer that reduces parasitic absorption at near-infrared wavelengths. The combination of these innovations resulted in a certified power conversion efficiency of 29.80%, demonstrating the significant potential of nano-optical engineering approaches.[69]

**Advanced Light Management and Photonic Structures**

Recent developments in light management techniques have focused on addressing both efficiency and stability challenges simultaneously. Haque et al. introduced innovative photonic structures that combine light trapping with UV protection capabilities. Their research presents a checkerboard (CB) tile pattern with designated UV photon conversion capability, representing a novel approach to addressing multiple performance-limiting factors in perovskite solar cells.

Through a combined optical and electrical modeling approach, the research demonstrates that this photonic structure can increase photocurrent and power conversion efficiency in ultrathin perovskite solar cells by 25.9% and 28.2%, respectively. These improvements are particularly significant for ultrathin devices, where optical losses typically limit performance.

The study further introduces a luminescent downshifting encapsulant that converts UV irradiation into visible photons matching the solar cell absorption spectrum. This approach addresses one of the critical stability challenges in perovskite solar cells, as UV radiation is known to cause degradation of perovskite materials. The research demonstrates that at least 94% of the impinging UV radiation can be effectively converted into the visible spectral range, providing both performance enhancement and stability improvement.

This work represents a significant advancement in addressing the dual challenges of efficiency and stability in perovskite solar cells. By combining light trapping with luminescent downshifting layers, the research unravels a potential photonic solution to overcome UV degradation while circumventing optical losses in ultrathin cells. This approach is particularly relevant for space applications under AM0 illumination, where UV radiation levels are significantly higher than terrestrial conditions.[73]

## Nanoscale Material Engineering and Enhancement Strategies

The strategic incorporation of nanoscale materials has emerged as a powerful approach for enhancing perovskite solar cell performance. Aftab et al. conducted a comprehensive investigation into the integration of various perovskite nanomaterials, including nanosheets (NSs), nanowires (NWs), nanorods (NRs), quantum dots (QDs), nanoparticles, and nanocrystals. Their research demonstrates that these nanoscale enhancements can significantly improve the optoelectronic properties and overall performance of perovskite-based photovoltaic devices.

The study reveals that the incorporation of nanoscale materials addresses several fundamental challenges in perovskite solar cells, including charge transport limitations, light absorption efficiency, and interface quality. The research demonstrates that different nanoscale morphologies offer unique advantages: nanosheets provide enhanced charge transport pathways, nanowires offer improved light scattering and absorption, while quantum dots enable tunable optical properties and enhanced stability.

The favorable optoelectronic properties and straightforward fabrication processes of perovskite-based solar cells, combined with nanoscale enhancements, have positioned these devices as promising contenders for next-generation photovoltaic technologies. The research emphasizes that the strategic selection and integration of appropriate nanomaterials can lead to synergistic effects that enhance multiple performance parameters simultaneously.[74]

## Comprehensive Progress in Efficiency and Stability

Recent comprehensive reviews have highlighted the remarkable progress achieved in perovskite solar cell technology across multiple dimensions. Wu et al. provided an extensive analysis of the main progress in perovskite solar cells during 2020-2021, focusing on efficiency improvements, long-term stability enhancements, and the development of perovskite-based tandem solar cells.

The review emphasizes that both efficiency and stability of perovskite solar cells have increased steadily in recent years, with particular attention to research aimed at reducing lead leakage and developing eco-friendly lead-free perovskites. This dual focus on performance and environmental considerations represents a crucial step toward the commercialization of perovskite solar cell technology.

The research highlights significant progress in perovskite-based tandem devices, which have emerged as one of the most promising pathways for achieving efficiencies beyond the single-junction limit. The development of these devices requires sophisticated interface engineering, optical optimization, and careful material selection to ensure compatibility between different cell components.

Furthermore, the review discusses the ongoing challenges and future prospects for perovskite solar cells, including the development of large-scale manufacturing processes, long-term stability under operational conditions, and the transition from laboratory-scale devices to commercial modules. The research emphasizes that addressing these challenges requires continued innovation in materials science, device engineering, and manufacturing technologies.[75]

## Emerging Applications and Future Directions

Beyond traditional photovoltaic applications, recent research has explored the versatility of perovskite materials in emerging technologies. Korde et al. investigated the application of perovskite nanostructures in biosensor applications, demonstrating the remarkable versatility of these materials beyond solar energy conversion. Their research reveals that the distinct presence of a central atom surrounded by eight ligands in perovskite structures leads to higher light absorption and charge carrier mobility, making these materials suitable for various sensing applications.

The capability of perovskite materials in detecting smaller molecules such as $O_2$, $NO_2$, and $CO_2$ has been demonstrated to be exceptionally high, leading to the development of several biosensors based on perovskite nanomaterials for detecting various chemical and biological species in both solid and solution states. This research highlights the broader potential of perovskite materials beyond photovoltaics and demonstrates their versatility as functional materials for diverse applications.

The exploration of perovskite materials in biosensor applications also provides insights that can be applied to photovoltaic devices. The understanding of charge transport mechanisms, surface interactions, and stability under various environmental conditions gained from biosensor research can inform the development of more robust and efficient solar cells.[76]

## Challenges and Future Prospects

Despite the remarkable progress achieved in recent years, several challenges remain to be addressed for the successful commercialization of perovskite solar cells. Afre and Pugliese provided a comprehensive assessment of the current state of the art in perovskite solar cell research, highlighting both achievements and remaining challenges.

The research emphasizes that while perovskite solar cells have demonstrated great potential as a low-cost alternative to conventional silicon-based solar cells, further research is required to improve their stability under ambient conditions for definitive commercialization. The mechanical stability of flexible perovskite solar cells represents another area of research that has gained significant attention, as flexible devices offer unique application opportunities but face additional stability challenges.

Recent research efforts have also focused on developing tin-based perovskite solar cells that can overcome the challenges associated with lead-based perovskites. These efforts address both environmental concerns and potential regulatory restrictions on lead containing materials. The development of lead-free alternatives represents a crucial step toward widespread adoption of perovskite solar cell technology.

The research also highlights the importance of developing novel materials for charging transport layers and advanced encapsulation techniques to protect perovskite solar cells from moisture and oxygen. These strategies are essential for achieving the long-term stability required for commercial applications.[77]

**Integration of Advanced Characterization Techniques**

The advancement of perovskite solar cell technology has been significantly aided by the development and application of sophisticated characterization techniques. The integration of advanced optical, electrical, and structural characterization methods has enabled researchers to gain deeper insights into the fundamental processes governing device performance and stability.

Recent research has demonstrated that the combination of multiple characterization techniques provides a comprehensive understanding of perovskite device operation. This multi-faceted approach has been crucial for identifying performance-limiting factors, optimizing device structures, and developing strategies for improving both efficiency and stability.

The application of in-situ characterization techniques has been particularly valuable for understanding degradation mechanisms and developing mitigation strategies. These techniques enable real-time monitoring of device performance under various stress conditions, providing insights that are essential for developing robust and stable perovskite solar cells. The continued development and refinement of characterization techniques will be crucial for the future advancement of perovskite solar cell technology as devices become more complex.[64]

**Comprehensive Analysis of High-Efficiency Perovskite Solar Cells**

Recent comprehensive reviews have provided detailed analyses of the current state and future prospects of high-efficiency perovskite solar cells. Dastgeer et al. conducted an extensive examination of the critical aspects that define high-efficiency perovskite solar cell technology, focusing on perovskite material properties, device configurations, fabrication techniques, and the latest advancements in the field. The research emphasizes that perovskite solar cells have achieved remarkable progress, with power conversion efficiencies reaching an astonishing peak value of 25.7%, demonstrating their disruptive potential in the photovoltaic sector. The study traces the evolution of inverted perovskite solar cells, which made their debut in 2013 with an efficiency of 3.9% and have since achieved significant milestones, including efficiencies exceeding 22.6% by 2021.

The comprehensive analysis addresses vital factors that are pertinent to the advancement of perovskite solar cell technology, including stability concerns, environmental impact, production scalability, and device reproducibility. The research highlights the ongoing challenges related to perovskite degradation and emphasizes the importance of developing improved stability protocols, enhancing efficiency, and integrating energy storage solutions to drive advancements in manufacturing.

Furthermore, the study discusses emerging trends in tandem and multijunction devices, flexible and wearable applications, and the integration of perovskite solar cells into building-integrated photovoltaic systems. The research provides insights into the commercialization pathway for inverted perovskite solar cells, underscoring the importance of stability, cost reduction, and efficiency enhancement in achieving widespread adoption of this promising technology.[78]

**Comparative Evaluation of Solar Cell Technologies**

A comprehensive evaluation of various solar cell technologies has provided valuable insights into the competitive landscape and relative advantages of perovskite solar cells. Oni et al. presented an in-depth assessment of cutting-edge solar cell technologies, emerging materials, loss mechanisms, and performance enhancement techniques across multiple photovoltaic platforms.

The study encompasses silicon and group III-V materials, lead halide perovskites, sustainable chalcogenides, organic photovoltaics, and dye-sensitized solar cells, providing a comprehensive comparison of their respective advantages and limitations. The research highlights that while the efficiency of silicon-based solar cells has plateaued around 25%, the efficiency of III-V compound semiconductor-based multijunction solar cells continues to increase, though the high material cost of III-V compound semiconductors remains a significant drawback.

The evaluation reveals that perovskite solar cells are extremely efficient in both single and multijunction arrangements, positioning them as strong competitors to established technologies. However, the research also identifies critical challenges that must be addressed for the industrialization of perovskite solar cells, including device deterioration, hysteresis effects, and film quality issues. The study emphasizes that CIGS and CdTe solar cell technologies compete with crystalline solar cells due to recent advances in cell performance, though environmental concerns and the low open-circuit voltage of CdTe solar cells remain significant challenges. This comparative analysis provides valuable context for understanding the position of perovskite solar cells within the broader photovoltaic technology landscape.[79]

**Roadmap for Perovskite Nanophotonics Applications**

The development of perovskite nanophotonics represents an emerging frontier that combines materials synthesis with novel photonic design strategies. Soci et al. presented a comprehensive roadmap for perovskite nanophotonics, outlining the current state and future directions of this rapidly evolving field.

The roadmap encompasses the integration of recent advances in synthetic material design, the development of bottom-up and top-down nanostructuring approaches, and new concepts in nanophotonic engineering of light-matter interaction at the nanoscale. This multidisciplinary approach has the potential to significantly impact current and future technologies by leveraging the unique properties of halide perovskite compounds.

The research identifies that perovskite nanophotonics are ready to blossom by combining pioneering works on halide perovskite compounds that have shown great potential across optoelectronics and photonics applications. The field represents a collective outlook from pioneers who have identified current and future challenges while highlighting the most promising research directions.

The roadmap serves as a comprehensive reference for physicists, chemists, and engineers interested in perovskite nanophotonics, providing guidance for future research and development efforts. The integration of nanophotonic concepts with perovskite materials opens new possibilities for advanced optical devices, enhanced light management in solar cells, and novel photonic applications.

This roadmap is particularly relevant for the development of advanced perovskite solar cells, as it provides insights into how nanophotonic engineering can be leveraged to improve light absorption, reduce optical losses, and enhance overall device performance. The combination of materials science and photonic design represents a promising pathway for achieving next-generation perovskite photovoltaic devices with superior efficiency and functionality.[70]

**Comprehensive Nanophotonic Design Strategies**

Recent comprehensive reviews have highlighted the transformative potential of nanophotonics in advancing perovskite solar cell technology. Furasova et al. provided an extensive analysis of nanophotonic design implementations for perovskite solar cell efficiency enhancement, covering critical methodologies including nanopatterning, nanotexturing, and nanostructuring approaches. Their work demonstrates how architectural modifications not only optimize optical properties but also significantly influence charge carrier transport and harvesting mechanisms. The review emphasizes the remarkable efficiency progression achieved through nanophotonic integration, with perovskite solar cells advancing from 3.8% efficiency in 2009 to 25.5% for single-junction devices and exceeding 30% for tandem configurations. Notably, the authors highlight the superior integration compatibility of nanophotonic designs with perovskite materials compared to silicon-based systems, attributed to the inherent flexibility and defect tolerance of halide perovskites.

The implementation of machine learning techniques for perovskite solar cell design optimization represents another significant advancement covered in this comprehensive review. These computational approaches enable systematic exploration of design parameters and accelerate the identification of optimal nanophotonic configurations. The work particularly emphasizes the potential for achieving efficiencies approaching the Shockley-Queisser limit of 30-35% through strategic nanophotonic design integration.[63]

**Advanced Photonic Design Principles for Next-Generation Photovoltaics**

Garnett et al. presented a broad perspective on photonic design opportunities across various photovoltaic technologies, with significant implications for perovskite solar cell development. Their analysis addresses critical photonic design principles including nanopatterning methods and metasurfaces for enhanced light incoupling and light trapping in absorber materials. The work identifies key opportunities for reducing carrier recombination through controlled light emission and explores advanced spectral conversion techniques that can be particularly beneficial for perovskite-based systems.

The perspective emphasizes the crucial role of photonic design in next-generation photovoltaic concepts, including tandem and self-adaptive solar cells where perovskites play a central role. The authors address the fundamental challenge of approaching the Shockley-Queisser efficiency limit of 34% through improved photonic design, highlighting how elimination of losses from incomplete absorption and nonradiative recombination can be achieved through strategic nanophotonic implementation. Their work provides a roadmap for massive upscaling and integration of photovoltaics, addressing both technical challenges and opportunities in photonic design principles and fabrication methodologies.[23]

## 2.2 Concentrating Solar Power with Light-Trapping Nanostructures

### 2.2.1 Overview of Concentrating Solar Power Technology

Concentrating Solar Power (CSP) represents a mature and rapidly evolving solar thermal technology that converts concentrated sunlight into thermal energy, which is subsequently converted to electricity through conventional thermodynamic cycles. Unlike photovoltaic systems that directly convert sunlight to electricity, CSP systems utilize mirrors or lenses to concentrate solar radiation onto receivers, where the concentrated thermal energy heats a working fluid to drive turbines for electricity generation.[80]

The fundamental principle of CSP technology lies in its ability to achieve high concentration ratios, typically ranging from 50-100 times for line-focusing systems to several thousand times for point-focusing systems. This concentration enables operation at elevated temperatures (500-800°C), resulting in higher thermodynamic efficiency and the unique capability to integrate thermal energy storage systems for dispatchable power generation.[81]

CSP technology encompasses four primary configurations: parabolic trough collectors (PTC), solar power towers, parabolic dish concentrators, and linear Fresnel reflectors. Each configuration employs different optical concentration strategies and operates at distinct temperature ranges, with solar towers and parabolic troughs leading the commercial market due to their scalability and proven performance.[82]

The critical component in all CSP systems is the solar receiver, where concentrated solar radiation is absorbed and converted to thermal energy. The receiver's performance directly impacts the overall system efficiency, making the development of advanced receiver technologies, particularly those incorporating nanophotonic principles, essential for next-generation CSP systems operating at temperatures exceeding 700°C.[83]

Nanophotonics has emerged as a transformative approach for enhancing CSP system performance through precise control of light-matter interactions at the nanoscale. The integration of nanophotonic structures enables the development of spectrally selective absorbers with tailored optical properties, achieving solar absorptance values exceeding 95% while maintaining infrared emissivity below 4% at elevated operating temperatures.

These engineered surfaces utilize photonic crystals, metamaterials, and plasmonic nanostructures to manipulate electromagnetic radiation across different spectral ranges, enabling optimal energy harvesting from concentrated sunlight while minimizing thermal radiation losses. The nanophotonic approach has proven particularly effective for CSP applications requiring high-temperature operation, where conventional selective coatings often suffer from thermal degradation and reduced spectral selectivity.[84]

### 2.2.2 Recent Developments in Concentrating Solar Power

**Femtosecond Laser Nanostructuring for Enhanced Solar Absorption**

Recent advances in ultrashort laser pulse technology have enabled the development of sophisticated nanostructured surfaces for CSP applications. Santagata et al. demonstrated significant improvements in solar absorptance through femtosecond laser texturing of commercial molybdenum surfaces. Their work achieved solar absorptance values approximately four times higher than pristine molybdenum samples at 800 K, with maximum improvements observed for surfaces treated at intermediate laser fluences (1.8 to 14 J/cm$^2$).

The enhancement mechanism relies on laser-induced periodic surface structures (LIPSS) that create subwavelength periodicity patterns on the molybdenum surface. These nanostructures function as light-trapping elements, significantly increasing the effective absorption area and surface roughness. The study revealed that LIPSS formation results from interference between incident laser waves and surface plasmon polaritons, leading to periodic ripples with periodicities comparable to or smaller than the laser wavelength.

Thermal stability testing demonstrated that the nanostructured molybdenum surfaces maintain their enhanced optical properties even after prolonged thermal annealing at operating temperatures typical of thermionic converters. The research established that intermediate laser fluences provide optimal performance by balancing enhanced light trapping with controlled oxide formation, while excessive fluences lead to significant oxide presence that reduces selectivity despite maintaining high absorptance.[80]

**Multi-Scale Light-Trapping Nanostructured Coatings**

The development of light-trapping nanostructured coatings represents a breakthrough approach for next-generation CSP receivers operating at elevated temperatures. Wang et al. developed a comprehensive multi-scale modeling framework combining Monte Carlo Ray Tracing (MCRT), Finite Difference Time Domain (FDTD), and Finite Volume Method (FVM) to evaluate receiver performance across nine orders of magnitude, from heliostat fields (~10 m) to nanostructured coatings (~100 nm).

Their investigation of three distinct nanostructure geometries—pyramid, moth-eye, and cone structures—revealed that cone nanostructures achieve superior optical-thermal performance with receiver efficiencies exceeding 88%. This represents a 6-10 percentage point improvement over commercial Pyromark 2500 coatings. The enhanced performance stems from optimized light-trapping mechanisms that increase solar absorption while maintaining low thermal emittance.

The multi-scale modeling approach successfully bridged the gap between nanoscale optical phenomena and system-level performance, enabling accurate prediction of how nanostructured coatings influence overall CSP plant efficiency. The study demonstrated that metallic nickel-based nanostructured arrays provide intrinsically low emissivity while achieving high solar absorption through carefully engineered surface texturing.[81]

**High-Performance Multilayer Selective Solar Absorbers**

Advanced multilayer selective solar absorber configurations have emerged as promising solutions for high-temperature CSP applications. Farchado et al. developed a novel six-layer selective absorber ($SiO_2/PtAl_2O_3/Pt/PtAl_2O_3/CuCoMnOx/SiO_2$) designed for operation at temperatures up to 550°C in air atmosphere, addressing the critical challenge of oxidation-induced degradation in next-generation CSP systems.

The optimized multilayer structure achieved exceptional optical performance with solar absorptance of 0.957 and thermal emittance of 0.10 at 500°C when deposited on stainless steel 316L substrates using cost-effective dip-coating methodology. The design philosophy incorporates platinum-based intermediate layers that provide oxidation resistance while maintaining high optical performance, and copper-cobalt-manganese oxide (CuCoMnOx) as the primary absorbing layer.

Durability testing demonstrated outstanding thermal stability, with the multilayer stack withstanding 3072 hours at 500°C in open air atmosphere without performance degradation (PC < 0.01). The material's resistance to condensation conditions and vacuum loss makes it particularly suitable for parabolic trough receivers, where maintaining performance during operational transients is critical. XRD analysis confirmed high crystallinity of constituent layers, while XPS depth profiling verified the integrity of the six-layer sequence.[82]

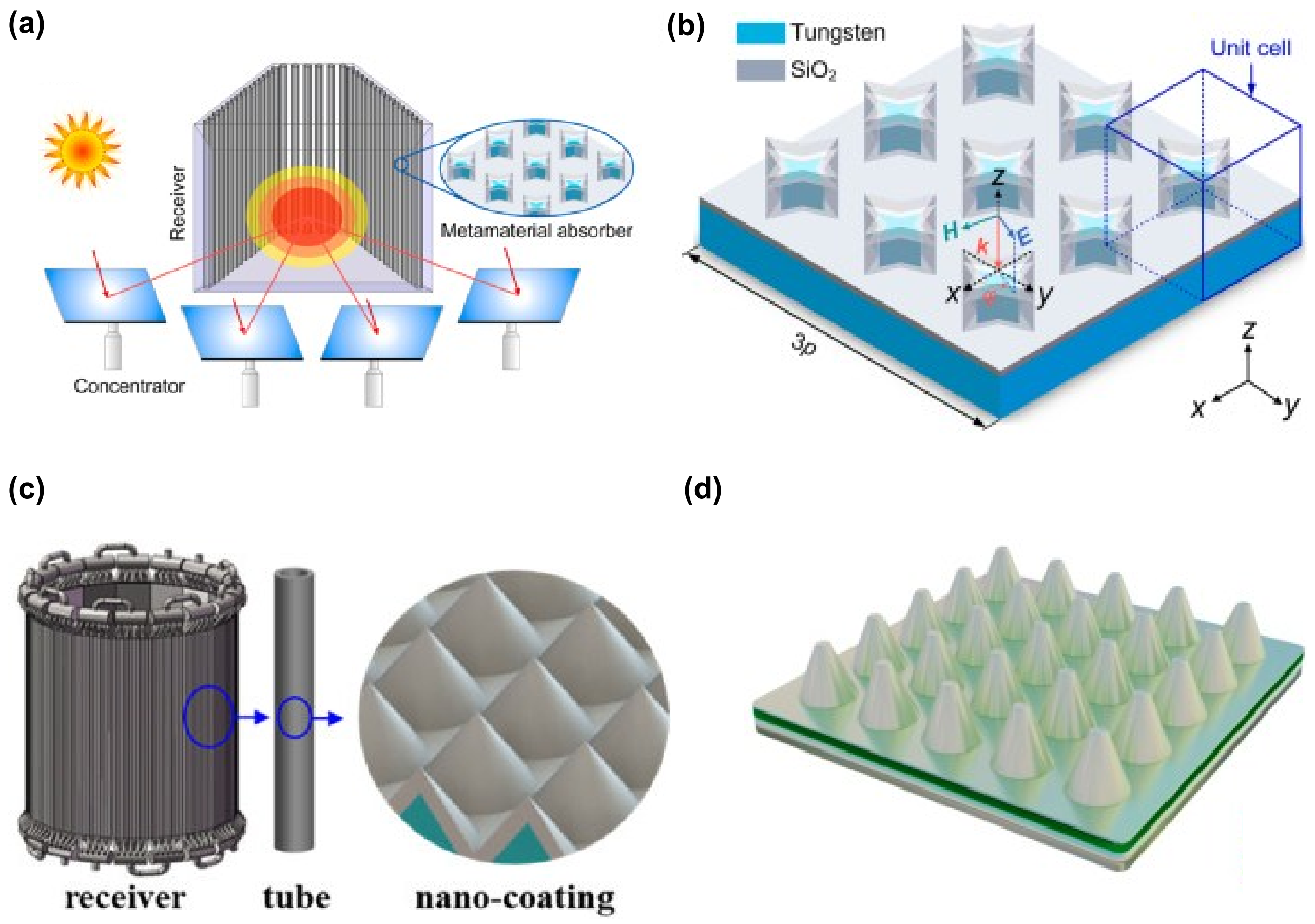

**Figure 4.** Some of the recent advancements in nanophotonics for enhancing the performance of concentrated solar power collectors. (a) solar receiver with a metamaterial absorber. (b) 3D periodic structural configuration of the receiver.[24] (c) schematic of a receiver with light-trapping nanostructured coatings. [81] (d) 3D schematic of the Meta-structure Solar Absorber (MSSA).[83]

### Plasmonic Meta-Structure Solar Absorbers

The development of plasmonic nanostructures for broadband solar absorption has achieved remarkable progress through synergistic combination of multiple absorption mechanisms. Su et al. designed a high-performance meta-structure solar absorber (MSSA) based on tungsten truncated cone structures integrated with metal-insulator-metal (MIM) film resonator configurations.

The optimized design achieved exceptional performance metrics including total solar absorption efficiency exceeding 97.1% and thermal emissivity below 8.5% under one solar concentration, resulting in 91.6% photothermal conversion efficiency at 100°C. The broadband absorption capability spans the ultraviolet, visible, and near-infrared regions (280-1700 nm) with absorptance exceeding 97.8%.

The enhanced absorption performance results from synergistic effects of multiple mechanisms: magnetic polaritons (MPs) on nanostructured tungsten surfaces, cavity plasmon resonance between truncated cones creating light-trapping structures, magnetic field resonance in the MIM optical resonator, and inherent tungsten losses. The impedance matching with free space ensures efficient light coupling, while the tungsten material provides exceptional thermal stability with melting point around 3400°C.[83]

### Four-Pointed Star Metamaterial Absorbers for High-Temperature Applications

Advanced metamaterial absorber designs have demonstrated exceptional potential for high-temperature solar energy harvesting applications. Qiu et al. developed a selective metamaterial absorber featuring four-pointed star prism structures optimized for operation at extreme temperatures up to 1673 K under concentrated solar conditions.

The optimized absorber achieved remarkable spectral selectivity with total solar absorptance of 0.958 and total emittance ranging from 0.2355 to 0.4062, enabling solar-to-heat conversion efficiencies of 92.31%-77.78% at 1000 suns concentration and temperatures of 1273-1673 K. The high spectral absorptance in the solar spectrum results from impedance matching and coupling effects of different plasmonic modes, while low spectral emittance in the mid-infrared band is achieved through impedance mismatching.

Parametric studies revealed that specific geometric parameters, including the height of the $SiO_2$ layer above the star structure and the short diagonal length of the rhomb, have minimal influence on spectral absorptance within certain ranges, providing design flexibility for manufacturing tolerances. The absorber demonstrates excellent insensitivity to polarization angle and incident angle variations, making it suitable for practical CSP applications where solar tracking precision may vary.

The research established that refractory metals like tungsten provide optimal characteristics for high-temperature metamaterial absorbers due to their high absorptance in the near-infrared wavelength range resulting from interband transitions, combined with exceptional thermal stability. These properties make four-pointed star metamaterial absorbers promising candidates for next-generation CSP systems operating at ultra-high temperatures.[24]

**One-Dimensional Multilayer Nanostructures for High-Temperature Applications**

Yuan et al. (2024) developed a high-temperature solar selective absorber based on simple one-dimensional multilayer nanostructures, addressing the critical challenge of balancing thermal performance with structural complexity in concentrated solar power systems. The research focused on creating a cost-effective solution that maintains excellent performance while reducing fabrication complexity compared to existing three-dimensional metamaterial structures.

The proposed absorber consists of alternating layers of tungsten (W) and silicon dioxide ($SiO_2$) in a one-dimensional multilayer configuration. The optimal design features ten layers with specific geometric parameters: $h_1$ = 90 nm, $h_2$ = 5 nm, $h_3$ = 80 nm, $h_4$ = 7 nm, $h_5$ = 80 nm, $h_6$ = 5 nm, $h_7$ = 60 nm, $h_8$ = 5 nm, $h_9$ = 60 nm, and $h_{10}$ = 200 nm. The material selection of tungsten (melting point 3422°C) and silicon dioxide (melting point 1723°C) ensures thermal stability under extreme operating conditions.[85]

The absorber demonstrates exceptional performance metrics with a total solar absorptance of 0.9504 to AM1.5 solar radiation. Under concentrated solar conditions of 1000 suns at 1273 K, the device achieves a remarkable solar-thermal efficiency of 91.2%, with a total emittance of only 0.1504 at the corresponding temperature. The structure exhibits high insensitivity to both incident polarization angles and wide-angle incidence, making it suitable for practical CSP applications where solar tracking precision may vary.

Comprehensive performance evaluation across concentration coefficients ranging from 300-1300 suns and temperatures from 500-1500 K confirmed the absorber's robust operation under diverse conditions. The spectral selective mechanism analysis revealed that the superior absorption performance is attributed to impedance matching properties resulting from the coupling effect of surface plasmon polaritons and magnetic polaritons.[85] This fundamental understanding provides insights for further optimization of one-dimensional nanostructured absorbers.

The fabrication advantages of this design are significant, as the one-dimensional multilayer structure can be manufactured using cost-effective magnetron sputtering coating techniques, in contrast to the expensive photolithography and micro-nano coating processes required for complex three-dimensional metamaterial structures. This approach substantially reduces manufacturing costs while maintaining high performance, making it more viable for large-scale CSP deployment.

## 2.3 Nanophotonics-based Solar Thermophotovoltaics

### 2.3.1 Overview of Solar Thermophotovoltaic Technology

Solar thermophotovoltaics (STPV) represents an innovative approach to solar energy conversion that has garnered significant attention due to its potential to exceed the Shockley-Queisser limit of conventional single-junction photovoltaic cells.[86] Unlike direct photovoltaic conversion, STPV systems employ a two-step energy conversion process: first converting concentrated solar radiation into thermal energy and subsequently converting thermal radiation from a heated emitter into electricity using photovoltaic cells.[87] This intermediate thermal step enables spectral control and optimization opportunities that can theoretically achieve conversion efficiencies exceeding 50%.[88]

The fundamental operating principle of STPV systems is governed by the Stefan-Boltzmann law and Planck's blackbody radiation theory.[89, 90] When concentrated solar radiation is absorbed by a thermal absorber, it heats the material to elevated temperatures, typically ranging from 1000 K to 2000 K. At these temperatures, the heated material acts as a thermal emitter, radiating electromagnetic energy according to Planck's distribution. The spectral characteristics of this thermal radiation can be tailored through careful engineering of the emitter's optical properties, enabling optimization for specific photovoltaic cell bandgaps.

The theoretical foundation for STPV efficiency stems from the limitations of conventional photovoltaic cells, which are fundamentally restricted by the Shockley-Queisser limit to approximately 31% efficiency under unconcentrated sunlight.[91] This limitation arises from the broadband nature of the solar spectrum, where sub-bandgap photons cannot be absorbed and above-bandgap photons result in thermalization losses. STPV systems address these limitations by absorbing sunlight across the entire solar spectrum and emitting narrowband infrared radiation matched to photovoltaic cell characteristics.

A typical STPV system consists of several critical components: a solar concentrator to focus incident sunlight, a solar absorber to convert concentrated solar radiation into thermal energy, a selective thermal emitter to control the spectral characteristics of emitted radiation, and photovoltaic cells optimized for the emitter's spectral output. Photovoltaic cells in STPV systems typically employ narrow-bandgap semiconductors such as gallium antimonide (GaSb), indium gallium arsenide (InGaAs), or indium gallium antimonide (InGaSb) to efficiently convert near-infrared thermal radiation.[92]

The role of nanophotonics in STPV systems has become increasingly critical as researchers seek to approach theoretical efficiency limits. Nanophotonic structures enable unprecedented control over thermal emission characteristics through engineered optical resonances, surface plasmon effects, and photonic bandgap phenomena. These capabilities allow the design of selective emitters with precisely tailored spectral properties, enhanced absorption and emission characteristics, and improved thermal stability. Photonic crystals, metamaterials, and metasurfaces have emerged as particularly promising approaches for achieving the spectral selectivity required for high-efficiency STPV operation.[93]

### 2.3.2 Recent Developments in Solar Thermophotovoltaics

The integration of nanophotonics with solar thermophotovoltaic systems has witnessed remarkable progress in recent years, with significant advances in selective emitter design, spectral control, and system efficiency. The following subsections detail the latest developments in STPV nanophotonics based on cutting-edge research from 2020-2025.

**Nanostructured Multilayer Selective Emitters**

Bhatt et al. demonstrated a high-efficiency STPV system employing a nanostructure-based selective emitter with multilayer metal-dielectric coatings.[94] The selective emitter consisted of $Si_3N_4$/W multilayer nanostructures deposited on a tungsten substrate, designed to achieve optimal spectral control for GaSb-based TPV cells. The experimental system achieved an electrical output power density of 1.71 W/cm$^2$ at an operating temperature of 1676 K, resulting in an overall power conversion efficiency of 8.4% when normalized to the emitter area. This represents the highest STPV efficiency reported at the time, demonstrating the effectiveness of nanostructured selective emitters in enhancing system performance. The multilayer design enabled precise control over the spectral emission characteristics, maximizing the overlap between the emitter spectrum and the TPV cell's spectral response while minimizing sub-bandgap losses.

**Comprehensive Nanostructure Design Strategies**

Gupta and Bhatt provided an extensive review of micro and nanostructure-based selective absorber and emitter surfaces for high-efficiency STPV applications.[93] Their work examined various nanostructure types including random textures, nanocones, nanoholes, and multilayer metal-dielectric stacks, demonstrating how these structures create interference effects for photons with wavelengths comparable to the feature sizes. The experimental validation using a $Si_3N_4$-W-$Si_3N_4$ selective emitter achieved an overall power conversion efficiency of 8.6% at 1670 K. The study emphasized that spectral selectivity through nanostructures enables STPV systems to surpass the Shockley-Queisser limit by tailoring the incident spectrum to match the TPV cell's bandgap, with theoretical efficiencies approaching 45% for optimized systems.

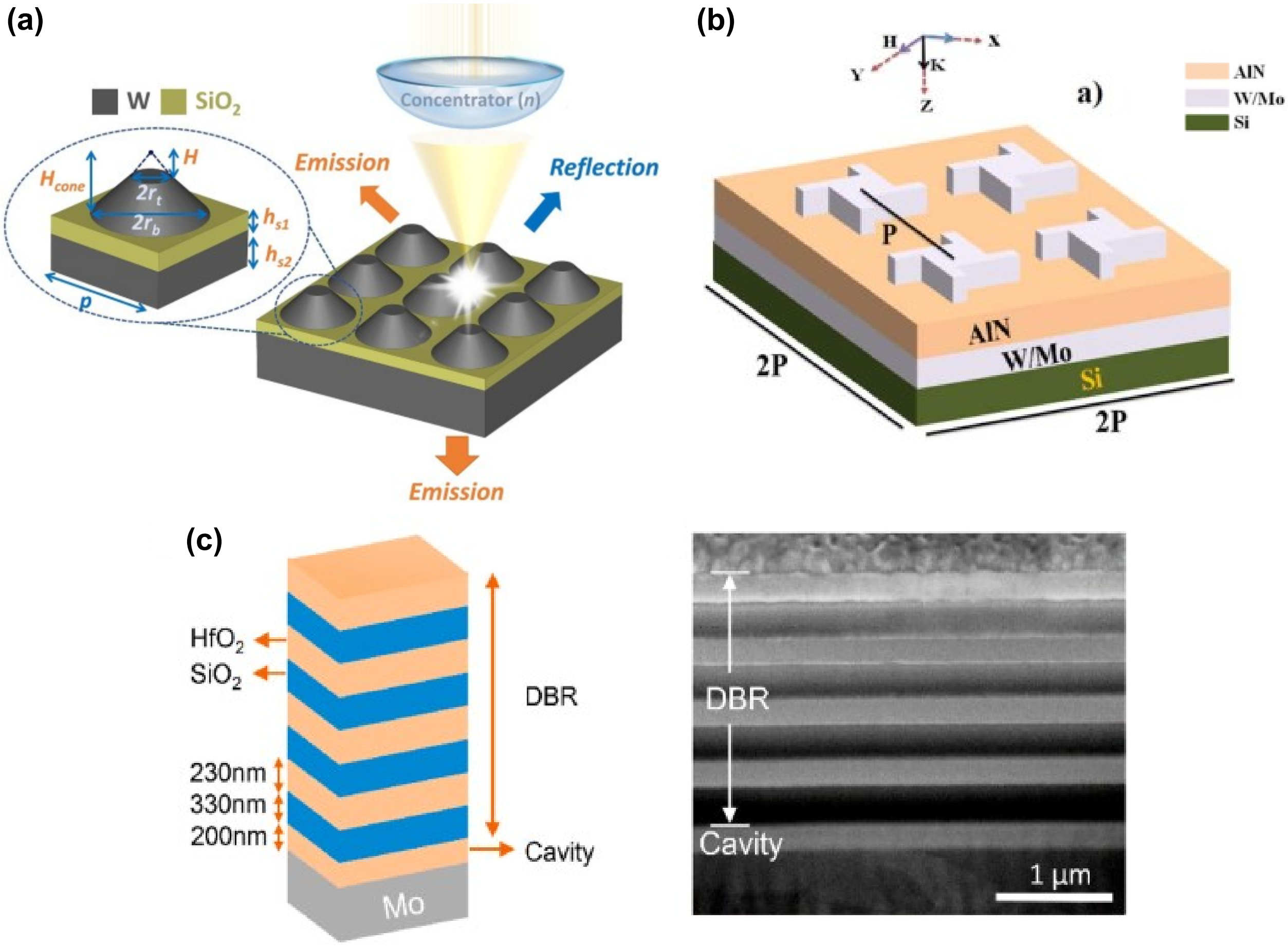

**Figure 5.** Some of the recent advancements in nanophotonics for enhancing solar thermophotovoltaic performance. (a) schematic of the $W/SiO_2/W$ structure featuring periodic truncated or full nanocones with labeled geometric parameters.[95] (b) three-dimensional nano-grating structure based on a metal-dielectric-metal (MDM) configuration.[25] (c) optical properties of a fabricated OTS-based emitter structure, including a multilayer design diagram with precise geometric details and a cross-sectional SEM image of the fabricated emitter.[96]

## Tamm Plasmon-Enabled Narrowband Thermal Emitters

Lin et al. developed a spectrally selective thermal emitter based on optical Tamm states (OTS) in a one-dimensional photonic structure.[96] The emitter consisted of alternately stacked $HfO_2$ and $SiO_2$ layers on a molybdenum substrate, achieving a maximal simulated thermal emissivity of 0.97 with an exceptionally narrow bandwidth of 48 nm at 1.9 μm wavelength. The overall system efficiency of the proposed emitter-based STPV device reached 33.7% under solar concentration of 2500 and an area ratio of 20. The optical Tamm state mechanism enabled precise spectral control by confining electromagnetic fields at the interface between the metallic substrate and the photonic crystal structure, demonstrating the potential of quantum optical phenomena for advanced STPV applications.

**Nanolayered Wavelength-Selective Emitters**

Wang et al. presented nanolayered narrowband thermal emitters utilizing Tamm plasmon polaritons (TPPs) in a-$SiN_x$ and a-$SiN_\gamma O_\gamma$ alternatively stacked nanolayers.[97] The emitters were fabricated on polished silicon substrates covered with metallic molybdenum, exhibiting narrowband absorption with absorptance above 90% at the designed emission wavelength. The simulated STPV system efficiency reached 28.9% under solar concentration of 1000. A key advantage of the molybdenum-based design was the suppressed absorption as low as 1.4% in the 2-7 μm wavelength region, combined with high resistance to high-temperature treatment in air atmosphere. The tunability of absorption spectra through simple thickness variation of the multilayers provided design flexibility for different TPV cell bandgaps.

**Dual-Coherence Enhanced Absorption Systems**

Zhang et al. introduced a novel spectrum-selective high-temperature tolerant thermal emitter based on dual-coherence enhanced absorption (DCEA).[98] The design employed a few-layer Si/Mo/AlN lamellar film on a molybdenum substrate, fabricated through sequential physical vapor deposition. The emitter exhibited maximal emissivity of approximately 97% at 1.4 μm with excellent thermal suppression down to 10% across the 3.4-10 μm range, maintaining performance stability up to 973 K. When implemented in STPV systems, the DCEA approach contributed to a 20% increase in total system efficiency compared to unimodal coherent perfect absorption counterparts, demonstrating the advantages of multi-path optical interference for enhanced spectral control.

**Metamaterial Selective Emitters for High-Bandgap Cells**

Tian et al. designed a metamaterial selective emitter based on tantalum and $Al_2O_3$ specifically optimized for high-bandgap silicon cells.[99] The concentrated STPV system utilizing the cavity-structured absorber and metamaterial selective emitter achieved an efficiency of 37.18%. A significant finding was that the available proportion of spectrum for Si cells increased dramatically from 16.52% to 72.69% through spectral reshaping using the designed selective emitter. The study demonstrated that STPV systems using Si cells achieved 12-13 percentage points higher efficiency compared to systems using low-bandgap cells such as GaSb, highlighting the potential of high-bandgap approaches when combined with appropriate nanophotonic spectral control.

**Non-Hermitian Selective Thermal Emitters**

Prasad and Naik developed a groundbreaking approach using non-Hermitian optics for selective thermal emission.[100] Their hybrid metal-dielectric non-Hermitian selective emitter (NHE) achieved high spectral efficiency exceeding 60% and demonstrated a maximum TPV conversion efficiency of 12% at 1273 K. The non-Hermitian approach represents a novel quantum-inspired methodology for controlling thermal emission, offering new possibilities for enhancing spectral selectivity beyond conventional approaches. The work addressed fundamental limitations of traditional selective emitters by leveraging non-Hermitian optical phenomena to achieve superior performance characteristics.

**Nanoscale Grating Metamaterial Emitters for High-Temperature Applications**

Feyisa et al. developed nanoscale grating metamaterial emitters based on tungsten/molybdenum films with aluminum nitride spacers for thermophotovoltaic applications.[25] The three-dimensional metal-dielectric-metal (MDM) grating structures achieved exceptional performance with 94% mean emittance for W-AlN-W structure in the wavelength range of 0.3-2.2 μm and 93% for Mo-AlN-Mo structure in the 0.3-2.0 μm range at normal incidence. The optimized emitter design demonstrated high spectral efficiency of 87% and 87.5% at 1600K for InGaAs bandgaps of 0.55 eV and 0.62 eV, respectively. The metamaterial emitters exhibited polarization independence and maintained good emissivity over a wide range of incidence angles from 0° to 75°. The high performance was attributed to the synergistic effects of surface plasmon polaritons, magnetic polaritons, and intrinsic metal absorption mechanisms. The design offers advantages of high thermal stability, easy fabrication, cost-effectiveness, and the unique capability of using one structure for two different bandgaps.

**Near-Field Thermophotovoltaic Devices**

Song et al. investigated the effectiveness of multi-junction cells in near-field thermophotovoltaic (NF-TPV) devices, focusing on photon tunneling effects and scalability considerations.[101] Their comprehensive study examined the impact of additional losses arising from high photocurrent densities and developed approximative expressions to quantify these effects. The research provided crucial insights into the scalable design of NF-TPV devices, emphasizing the role of multi-junction PV cells in enhancing power output density through near-field radiative heat transfer enhancement. The analytical framework developed enables precise performance estimations for devices with 10 or more subcells, facilitating optimization across various parameters including vacuum gap distance and emitter temperature.

**Metasurface-Controlled Thermal Emission**

Chu et al. provided a comprehensive review of controlling thermal emission with metasurfaces and their applications in thermophotovoltaic systems.[102] The review highlighted recent advances in tuning thermal emission across multiple degrees of freedom including wavelength, polarization, radiation angle, and coherence using two-dimensional subwavelength artificial nanostructures. Metasurfaces offer unprecedented flexibility in shaping thermal emission characteristics, enabling compact and integrated optical devices for STPV applications. The work emphasized the transition from broadband, unpolarized, and incoherent conventional thermal emission to precisely controlled emission through metasurface engineering.

**Chromium Metasurface Broadband Absorbers**

Rana et al. developed a broadband metasurface solar absorber composed of refractory chromium for STPV intermediate structures.[103] The metasurface absorber exhibited high broadband absorptance with an average exceeding 90% across the 300-1200 nm wavelength range. The proposed STPV system achieved a PV cell efficiency of 43.2% with efficiency greater than 42% maintained across a broad color temperature range of 1597-2573 K. The chromium material provided self-passivation properties ensuring resistance to oxidation and corrosion, combined with low cost and stability at high temperatures. The innovation stemmed from efficiency enhancement through hybridization of selectivity and broadband response of the absorbers and emitters.

**Nanocone-Based Photonic Crystal Absorbers**

Mirnaziry et al. investigated the optical and thermal response of 2D photonic crystal absorbers composed of tungsten nanocones for STPV applications.[95] The study examined both complete and truncated nanocone shapes, analyzing their thermo-optical performance through comprehensive modeling. The nanocone-based design leveraged photonic crystal effects to enhance solar absorption while maintaining thermal stability at high operating temperatures. The research provided insights into optimizing nanocone geometries for maximum absorption efficiency and thermal management in practical STPV systems.

**Optically Transparent Metasurface-Based STPV Systems**

Shafique et al. introduced a groundbreaking optically transparent metasurface (OTM) based STPV system that combines visible transparency with efficient solar energy conversion.[104] The innovative design employs indium tin oxide (ITO) as the transparent conducting metal and zinc sulfide (ZnS) as the substrate layer, creating a four-layer structure with cross-shaped resonators. The OTM structure demonstrates exceptional broadband absorption capabilities, achieving up to 99% absorption in the UV region (250-400 nm) and over 90% absorptivity in the far infrared range (800-2000 nm) while maintaining high average transmittance in the visible spectrum. This unique combination enables the system to harvest solar energy without compromising the transparency required for building-integrated photovoltaic (BIPV) applications. The design exhibits remarkable angular stability, maintaining absorption rates exceeding 90% even at incident angles up to 70° for both TE and TM polarizations. The transparent nature of the metasurface opens new possibilities for integration into building windows, vehicle surfaces, and portable electronic devices, potentially meeting over 40% of building energy consumption when applied to glass exteriors. The research demonstrates that transparent conducting oxides like ITO can serve as effective plasmonic materials for STPV applications, offering a cost-effective and scalable solution for large-scale fabrication while addressing the growing demand for aesthetically compatible renewable energy solutions.

# 3 Nanophotonics for Biosensing Applications

Nanophotonics has emerged as a transformative field in biosensing, offering unprecedented sensitivity, real-time operation, miniaturization, and multiplexing capabilities addressing critical gaps in diagnostics, environmental monitoring, and food safety. By leveraging light-matter interactions at the nanoscale, these biosensors enable rapid, label-free detection of biomolecules, pathogens, and environmental contaminants. This review synthesizes insights from 40+ recent studies (2018–2024) to examine the key Material innovations (plasmonic metals, 2D materials, hybrids) and their performance trade-offs, advantages over conventional techniques (e.g., ELISA, PCR), applications across clinical, environmental, and industrial sectors, challenges (fabrication, scalability) and future directions (AI integration, wearables).

## 3.1 Key Nanophotonic Materials for Biosensors

The performance of nanophotonic biosensors is heavily influenced by the choice of materials, which determine sensitivity, stability, and compatibility with biological samples. These materials are categorized as follows: 1) Plasmonic metals, such as gold (Au) and silver (Ag), are widely used due to their strong localized surface plasmon resonance (LSPR), which enhances electromagnetic fields for ultrasensitive detection. For example, Au-Ag bimetallic nanoparticles (NPs) amplify surface-enhanced Raman scattering (SERS) signals by 14fold, enabling single molecule detection of viruses [105 106], while Ag nanocubes show 10× higher SERS signals than spheres but require $SiO_2$ coatings to prevent oxidation [107]. Aluminum (Al) extends plasmonics into the ultraviolet range, facilitating applications like H. pylori detection in clinical samples [108]. 2) Dielectric materials, including silicon nitride ($Si_3N_4$) and titanium dioxide ($TiO_2$), are prized for their low optical losses and compatibility with complementary metal-oxide-semiconductor (CMOS) technology. $Si_3N_4$ waveguides, for instance, achieve refractive index resolutions of $10^{-7}$–$10^{-8}$ RIU, making them ideal for detecting SARS-CoV-2 at 200 copies/μL without amplification [109]. $TiO_2$ nanosheets, on the other hand, Combines sensing with photocatalytic pathogen inactivation, adding a therapeutic dimension to biosensing [110]. 3) Two-dimensional (2D) materials, such as molybdenum disulfide ($MoS_2$) and graphene, offer unique electronic and optical properties. $MoS_2$-based field-effect transistors (FETs) detect miRNA155 at concentrations as low as 0.03 fM, while graphene oxide aptasensors identify aflatoxin $B_1$ in food at 0.25 ng/mL, surpassing traditional ELISA methods [111 112]. 4) Carbon-based nanomaterials, including carbon quantum dots (CQDs) and carbon nanotubes (CNTs), provide biocompatibility and versatility. Nitrogen-doped CQDs synthesized from radish leaves detect the anticancer drug nintedanib at 0.14 μg/mL via fluorescence quenching, showcasing their potential for environmental monitoring [113]. CNTs, with their high aspect ratio and conductivity, are employed in electrochemical biosensors for real-time glucose monitoring in urine [114]. 5) Hybrid materials, such as metalorganic frameworks (MOFs) and core-shell nanoparticles, combine the strengths of multiple components. ZIF8 MOFs stabilize plasmonic gold nanorods (Au-NRs), enabling glyphosate detection in water, while silica-coated silver nanoparticles (Ag@$SiO_2$) reduce quenching in SERS tags, improving signal reproducibility [105 115]. 6) Emerging materials, like rare-earth-doped NPs and chiral meta-surfaces, address niche applications. Rare-earth elements (e.g., europium, dysprosium) enhance electrochemical biosensors for neurotransmitters, while chiral meta-surfaces enable enantiomer-specific detection of biomolecules, critical for pharmaceutical applications [116 117].

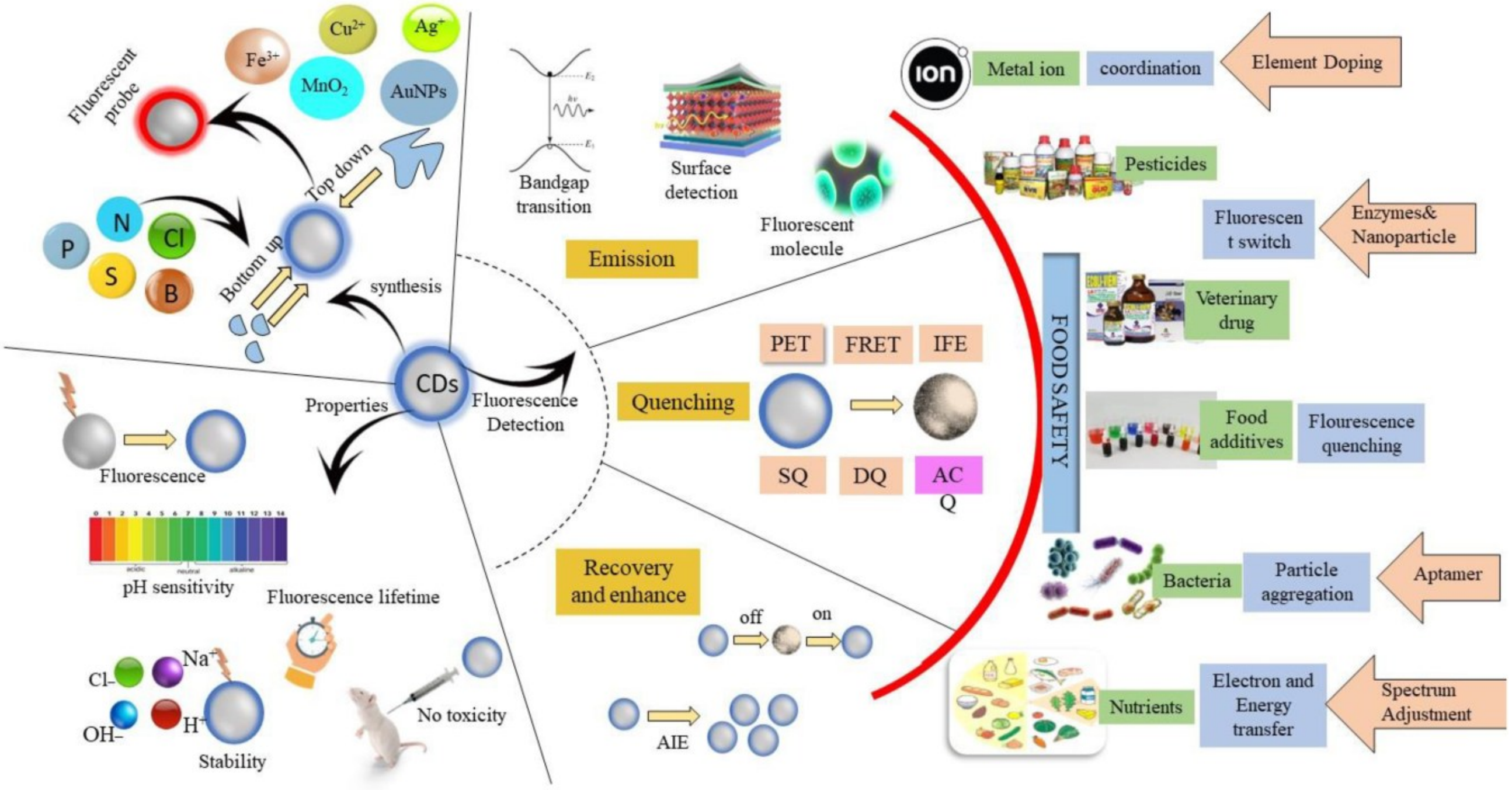

**Figure 6.** Nanosensors: A visual exploration of revolutionary applications. Reprinted and permission from Ref. [118]

## 3.2 Advantages of Nanophotonics in Biosensing

Nanophotonic biosensors offer several advantages over conventional techniques, including: 1) Ultrahigh sensitivity is a hallmark of nanophotonics, with LSPR biosensors detecting cardiac troponin I (cTnI) at 0.01 ng/mL and SERS achieving single molecule resolution [107 119], . Photonic crystal sensors further push the limits, resolving interleukin6 (IL6) at 10 pg/mL in wound exudate, enabling early diagnosis of chronic infections [120]. 2) Rapid and label-free operation eliminates the need for time-consuming sample preparation and fluorescent labeling. For example, silicon nitride ($Si_3N_4$) interferometers quantify whole SARSCoV2 viruses in under 20 minutes, rivaling PCR accuracy without amplification [109]. Similarly, plasmonic gold nanoparticles (AuNPs) provide real-time monitoring of thrombin/DNA interactions, crucial for studying coagulation disorders [111]. 3) Multiplexing allows simultaneous detection of multiple analytes, a critical feature for complex diagnostics. Quantum dot barcodes distinguish nine respiratory viruses in a single assay, while TriPlex™ photonic waveguides monitor pollutants like bisphenol A (BPA) and atrazine in water with limits of detection (LODs) as low as 0.06 µg/L [121] [122].

Furthermore, portability and integration with point-of-care (POC) devices are facilitated by miniaturization. Smartphone-coupled AuNP assays detect pesticides in food with visual readouts, and 3Dprinted graphene electrodes enable decentralized glucose monitoring [114] [123]. While, artificial intelligence (AI) integration is an emerging trend, enhancing data analysis and sensor performance. Deep learning models, such as convolutional neural networks (CNNs), reduce SERS spectral artifacts by 82%, enabling accurate classification of lung cancer biomarkers [111] [124].

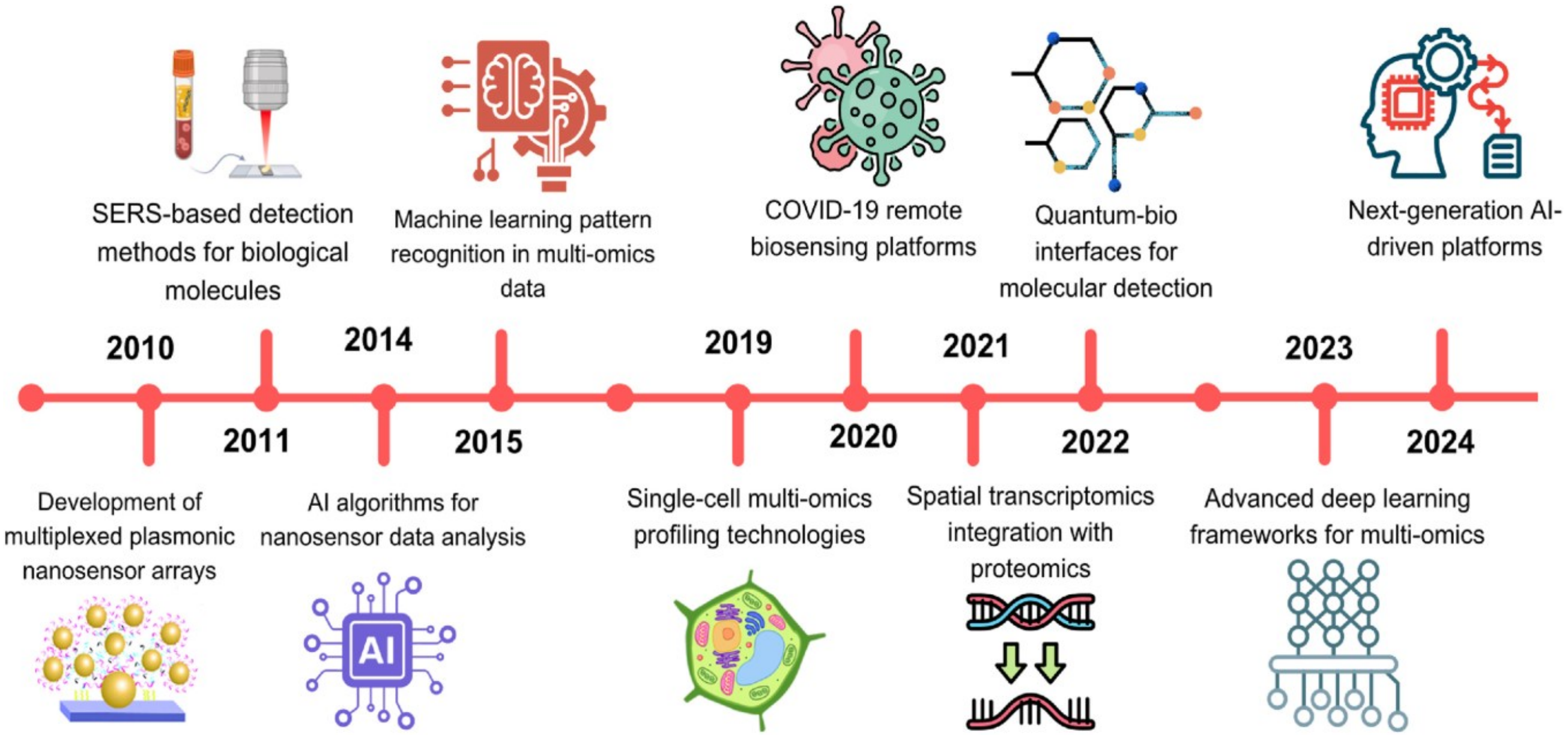

**Figure 7.** The timeline highlights key technological milestones in the convergence of optical nanosensors with multi-omics platforms, from early genomic-proteomic applications (2010–2015) to advanced AI-driven systems capable of single-cell resolution and real-time multimodal analysis (2020–2024). Critical developments include quantum-bio interfaces, edge computing solutions, and nanomaterial-enhanced AI frameworks. Adapted from Ref. [125] with permission from Elsevier.

## 3.3 Applications of Nanophotonic Biosensors

Nanophotonic biosensors have transcended laboratory research to become pivotal tools in real-world scenarios, revolutionizing detection capabilities across multiple disciplines. Their unparalleled sensitivity, multiplexing capacity, and miniaturized form factors enable transformative applications in precision medicine, global health surveillance, environmental protection, and food security.

### 3.3.1 Biomolecule Detection

These biosensors excel in identifying proteins, nucleic acids, and small molecules, which is critical for: 1) Early Disease Diagnosis: Detection of low-abundance biomarkers (e.g., cardiac troponin I at 0.01 ng/mL [119]) enables timely intervention for conditions like myocardial infarction. 2) Epigenetic Research: FRET-based quantum dots (QDs) map DNA methylation patterns (e.g., hypermethylation of tumor suppressor genes [125]) with 10× higher resolution than microarrays, advancing personalized cancer therapies. 3) Point-of-Care Metabolic Monitoring: Catalytic MOFs (e.g., Ce-MOFs [115]) provide continuous glucose tracking without finger-prick blood sampling, addressing a key need for 463 million diabetics worldwide. While Traditional techniques (e.g., Western blotting) require 24+ hours and lose sensitivity for rare biomarkers—a gap nanophotonics closes.

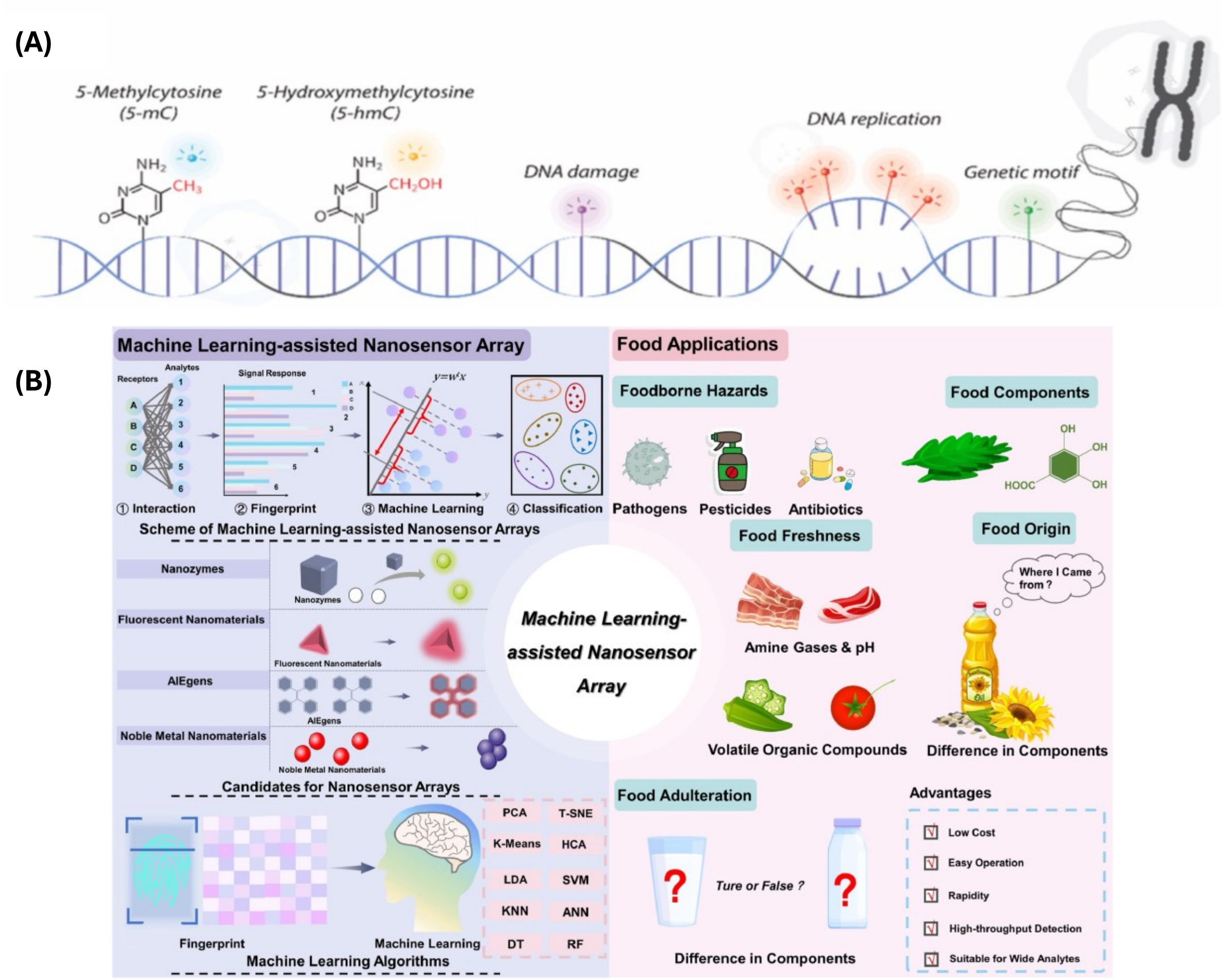

**Figure 8.** Example of optical nanosensor applications: (A) Epigenetic mapping of single DNA molecules - Reprinted and permission from Ref. [125], (B) Scheme of machine learning-assisted nanosensor arrays and their applications in food quality detection analysis -Reprinted and permission from Ref. [126]

### 3.3.2 Medical Diagnostics

Nanophotonic biosensors are transforming modern medicine by enabling earlier, more accurate disease detection than conventional methods. In infectious disease detection, the PANTOMIM biosensor identifies urinary tract infections (UTIs) at 1.23 CFU/mL, and SERS molecular beacons diagnose methicillin-resistant *Staphylococcus aureus* (MRSA) in 45 minutes [105] [127]. For cancer, Au-$MoS_2$ nanosheets detect carcinoembryonic antigen (CEA) at 1.6 fg/mL [112] - equivalent to finding 1 cancerous cell among $10^9$ healthy ones - potentially reducing late-stage cancer mortality by 50%.enabling early intervention. Chronic disease management benefits from guided-mode resonance (GMR) sensors that track IL6 in chronic wounds, predicting healing trajectories [120].

### 3.3.3 Environmental Monitoring

The unprecedented sensitivity of nanophotonic biosensors is addressing critical gaps in environmental surveillance. In mitigating water pollution with heavy metal toxicity, water quality assessment employs photonic crystal fibers to detect cadmium ($Cd^{2+}$) at $3.8 \times 10^{-11}$ M, while hyperspectral imaging identifies airborne pathogens in real time.[110, 112]

### 3.3.4 Food Safety

Nanophotonic biosensors are becoming essential tools for ensuring food safety from production to consumption. Addressing pathogen detection, toxin monitoring and supply chain integrity. AuNP-based lateral flow assays (LFAs) screen for E. coli O157:H7 in meat, and SERS nanotags quantify ochratoxin A in grains, ensuring compliance with safety standards.[128, 129]

## 3.4 Challenges and Future Directions

Despite their promise, nanophotonic biosensors face several challenges such as 1) Fabrication Complexity remains the foremost barrier to scalable deployment. While SERS substrates achieve remarkable sensitivity through precisely engineered nanostructures (e.g., 2 nm gap Au dimers [107]), their production relies on costly e-beam lithography (~$500/chip [130]) with <60% yield. Similarly, MOF synthesis often requires 72-hour solvothermal processes at 120°C [115], consuming 30× more energy than nanoparticle production. These limitations restrict applications where disposable, mass-produced sensors are needed, such as pandemic screening or home testing. 2) Matrix Interference severely compromises performance in real biological samples. Studies show serum proteins reduce LSPR shift signals by 40-60% [108], while stool samples introduce >80% false positives in pathogen detection [131]. Current solutions like centrifugal filtration add complexity, increasing processing time from minutes to hours—negating the technology's rapid response advantage. The lack of universal antifouling coatings forces case-by-case optimization, delaying clinical translation. 3) Scalability and cost remain hurdles, particularly for noble metal-based sensors and clean-room-fabricated photonic devices [130], [109]. 4) Standardization is lacking, with no universal protocols for performance validation [127] [107].

Future directions may include: 1) Hybrid Material Systems which show exceptional promise for overcoming stability-cost trade-offs for example: MOF-encapsulated plasmonic NPs (e.g., ZIF-8@AuNRs [115]) combine 90-day environmental stability with 0.1 nM sensitivity, while 2D material heterostructures (graphene-hBN-$MoS_2$ [112]) enable >100× reuse cycles while maintaining 90% initial sensitivity, also Bioinspired designs (mussel foot protein coatings [108]) reduce biofouling by 75% without sample pretreatment. 2) Wearable Integration is advancing through as flexible hydrogel patches with embedded GMR sensors [120] that monitor chronic wound biomarkers (IL-6, TNF-α) at 6-hour intervals, and cContact lens platforms using glucose-responsive photonic crystals (5-min refresh rate [114]), furthermore subdermal implants for continuous drug monitoring (phase I trials for vancomycin detection [128]). 3) AI/ML Co-Design is revolutionizing two fronts: 1) Fabrication Optimization: Generative adversarial networks (GANs) predict optimal nanostructures (e.g., 97% accuracy for SERS hot-spot design [124]), 2) Data Interpretation: Convolutional neural networks decode multiplexed SERS spectra with 95% accuracy vs. 68% for PCA [111], enabling 10-plex detection from single samples.

In summary, Nanophotonic biosensors stand at a critical inflection point, transitioning from lab marvels to essential tools addressing global health and environmental challenges. Their unparalleled sensitivity (reaching single-molecule detection [105]) and multiplexing capacity (9-plex in <20 mins [122]) already outperform gold-standard techniques across diagnostics, environmental monitoring, and food safety applications. However, as this analysis reveals, achieving widespread adoption demands coordinated advances: 1) Manufacturing Innovation: Development of low-cost, high-yield nanofabrication methods to replace cleanroom dependence. 2) Regulatory Frameworks: Establishment of ISO standards for performance validation and quality control. 3) Interdisciplinary Collaboration: Convergence of materials science, AI, and clinical research to address matrix challenges.

The coming decade will likely witness nanophotonics enabling: Personalized Medicine: Implantable sensors for real-time therapeutic drug monitoring, Precision Agriculture: Field-deployable phytopathogen detection at 1/100 current costs, Pandemic Prevention: Airport-based pathogen screening with 90% sensitivity <5 mins. While challenges remain formidable, the coordinated efforts of academia, industry, and regulators can transform these technological triumphs into tangible societal benefits—ushering in an era where advanced diagnostics are as accessible as smartphone technology.

## 4. Nanophotonics in Medicine and Healthcare

Nanophotonics has become a field for innovation in medicine and healthcare. This permits us to design novel diagnostic and therapeutic tools that take advantage of nanomaterial characteristics and light-matter interactions. They are used in detecting diseases at early stages and developing non-invasive medical procedures.[132, 133] However, the clinical use of nanophotonics applications in healthcare faces some challenges. For example, biological barriers, safety and toxicity concerns with nanoparticles, systemic obstacles, and a complex regulatory environment are some of these difficulties. The need for safer nanoprobes with few adverse side effects is always growing. This section discusses recent advances in the field of nanophotonics applications in medicine and healthcare. For example, recent advances in Photothermal Therapy, applications for AI in Nanophotonic Healthcare Devices, The Role in Detecting MicroRNA Cancer Markers, Enhanced Chemotherapy and Imaging Modalities, and Image-Guided Surgery (IGS) Enhanced by Nanophotonics.

## 4.1. Nanophotonics for Photothermal Therapy of Tumors

Photothermal therapy (PTT) is used in apoptosis and tumor elimination through nanomaterials that absorb laser energy and convert it to localized heat.[134] It can target the tumor, reducing the damage to nearby healthy tissue. A prominent development in this area is the combination of PTT with other treatment modalities in order to generate synergistic anti-tumor effects. Nanostructures in PTT make it more effective in cancer therapy.[135] Their nanoscale size enables passive accumulation in tumor tissues.[136] The distinctive optical characteristics of different nanomaterials, notably their high absorption in the near-infrared (NIR) region (700-1700 nm), are crucial to their application in PTT. The use of surface-modified NPs allowed selective delivery and led to a precisely controlled increase in the local temperature.[137]

### 4.1.1. Nanomaterials as Photothermal Agents (PTAs)

**Noble Metal-Based Nanomaterials**

Noble metal nanoparticles have received a lot of attention in PTT due to their significant surface plasmon resonance and ability to absorb light at particular near-infrared wavelengths. As a result, they may be utilized as excellent photosensitizers to facilitate photothermal conversion and increase their efficiency. In that section, we focus on the use of noble metal nanoparticles, including gold, silver, platinum, and palladium, in the field of cancer therapy, proposing combination techniques including PTT.[138]

The five main types of Au NPs used for PTT that have garnered the most interest throughout the preclinical development or clinical trial stages are Au nano shells, Au nanorods, Au nanostars, Au nanocages, and Au nanospheres. while size affects cellular uptake, as Small NPs have improved cellular absorption. Shape affects cellular uptake as well. For instance, Au nanorods (NRs) exhibit reduced cellular absorption compared to other Au nanostructures due to their aspect ratio.[139] Gold nanostars, with their sharp tips, can induce larger hot spots, leading to higher PCE compared to nanospheres or nanorods.[140] Gold Nanoshells have been researched in clinical trials with published efficacy in prostate cancer, 94% effective when applied during the ablative surgery, and 87.5% negative biopsies one year later.[141] On the other hand, Spherical AuSHINs are frequently favored over gold nanorods due to their reduced toxicity and improved colloidal stability, which may require cytotoxic surfactants during production.[142] novel delivery methods such as nanostraw-assisted injection have been investigated, with results indicating approximately 10-fold higher internalization of gold shell-isolated nanoparticles (AuSHINs) and a 2-fold higher reduction in breast cancer cell viability compared to conventional incubation. A silver nanoprism has been shown to have considerable promise in photothermal treatment (PTT) due to its high surface plasmon resonance band in the near infrared area. However, its instability in physicochemical conditions and extreme toxicity limited its future use.[143]

Palladium nanoparticles have strong thermal and chemical stability, catalytic activity, and tunable optical response.[144] Palladium is a key component in the bimetallic nanoparticles (Ag-Pd NPs) stabilized by elm pod polysaccharide (EPP). Under near-infrared laser irradiation (808 nm), Pd-containing NPs reached 53.8°C vs. 43.5°C for Ag-only NPs.[145] Platinum is the first metal to be utilized as a cancer therapy. Platinum nanoparticles (PtNPs) have been reported to exhibit anticancer activity and the ability to enhance anti-tumor therapy.[146] For instance, PEG@Pt/DOX is an integrated system with both therapeutic and diagnostic capabilities. These capabilities enable the use of computed tomography (CT) imaging along with chemotherapy and photothermal treatments together synergistically.[147] Hyaluronic acid (HA)-modified platinum nanoparticles (PtNPs) presented significant cytotoxicity towards the aggressive MDA-MB-231 cell line in vitro and inhibited tumor growth in vivo using PTT. This suggests that the targeting of HA-mediated and tumor-penetrating nanosystems can reasonably improve therapeutic performance in vivo. However, these platinum-based drugs have shown toxic effects on the kidney, brain, nerve tissue, and bone marrow, leading to side effects.[148]

**Carbon-Based Nanomaterials**

Carbon-based nanomaterials (CBNs) have received a lot of interest as photothermal agents because of their distinct optical, thermal, and chemical characteristics, as well as their large surface area and biocompatibility.[149] Carbon Nanotubes (CNTs) show strong NIR absorption, high PCE, high thermal conductance, and potential for drug delivery. Single-walled CNTs (SWCNTs) and multi-walled CNTs (MWCNTs) are two types of CNTs that have a wide range of properties. SWNTs are graphitic helical molecules with excellent physical and mechanical properties. SWNTs have strong water solubility, minimal toxicity, and great biological stability. Adding polyethylene glycol (PEG) to SWNTs enhances blood circulation time. When Cy5.5 was added to SWNTs, it displayed Cy5.5-coupled SWNTs-mediated PTT, resulting in systemic tumor ablation in mice due to the absorption at 808 nm. Cy5.5 improved the therapeutic index of PTT.

Multi-walled carbon nanotubes (MWNTs), which range in diameter from a few nanometres to a few micrometres, have become popular options for biological imaging, photothermal tumour ablation, and tumour medication administration. MWNTs may produce localised heat upon NIR exposure, which might cause the tumour to thermally annihilate. Pathological analysis of the kidney, spleen, liver, and heart revealed that these modified CNTs were biocompatible and did not cause any harm to any of the organs. These nano-agents have enormous promise in clinical and therapeutic applications, despite the fact that their long-term safety concerns are currently relatively minor.[149]

## Graphene and its Derivatives:

Graphene and its derivatives are utilized in PTT for a variety of features like biocompatibility, easy production method, tunable surface performance, and greater water solubility.[150] their spectrum exhibits significant NIR absorption, high PCE, a large surface area, and drug loading capability. Carbon Dots (CDs) are Zero-dimensional nanocompounds known for their excellent optical properties, biocompatibility, low cost, and are fluorescent for imaging. CDs can produce synergistic regulatory cell death (RCD) pathways, such as necroptosis, using imaging-guided PTTs. Soya lecithin-coated red fluorescent carbon dots LRCDs are an example show improved bioavailability and therapeutic properties in breast cancer.

**Table 1.** Summary of nanomaterials used in PPT and its features.

| Nanomaterial Type | Key Properties for PTT | Examples |
|---|---|---|
| **Carbon Nanotubes (CNTs)** | Strong NIR absorption, high PCE, high thermal conductivity, and drug delivery potential | SWNTs<br>MWNTs |
| **Graphene & Derivatives** | Exceptional NIR absorption, high PCE, large surface area, and drug loading platform | Graphene Nanosheet<br>Graphene Oxide (GO)<br>Reduced Graphene Oxide (rGO)<br>Graphene Quantum Dots (GQDs) |
| **Carbon Dots (CDs)** | Exceptional optical properties, biocompatibility, cost-effectiveness, and fluorescence for imaging | LRCDs |

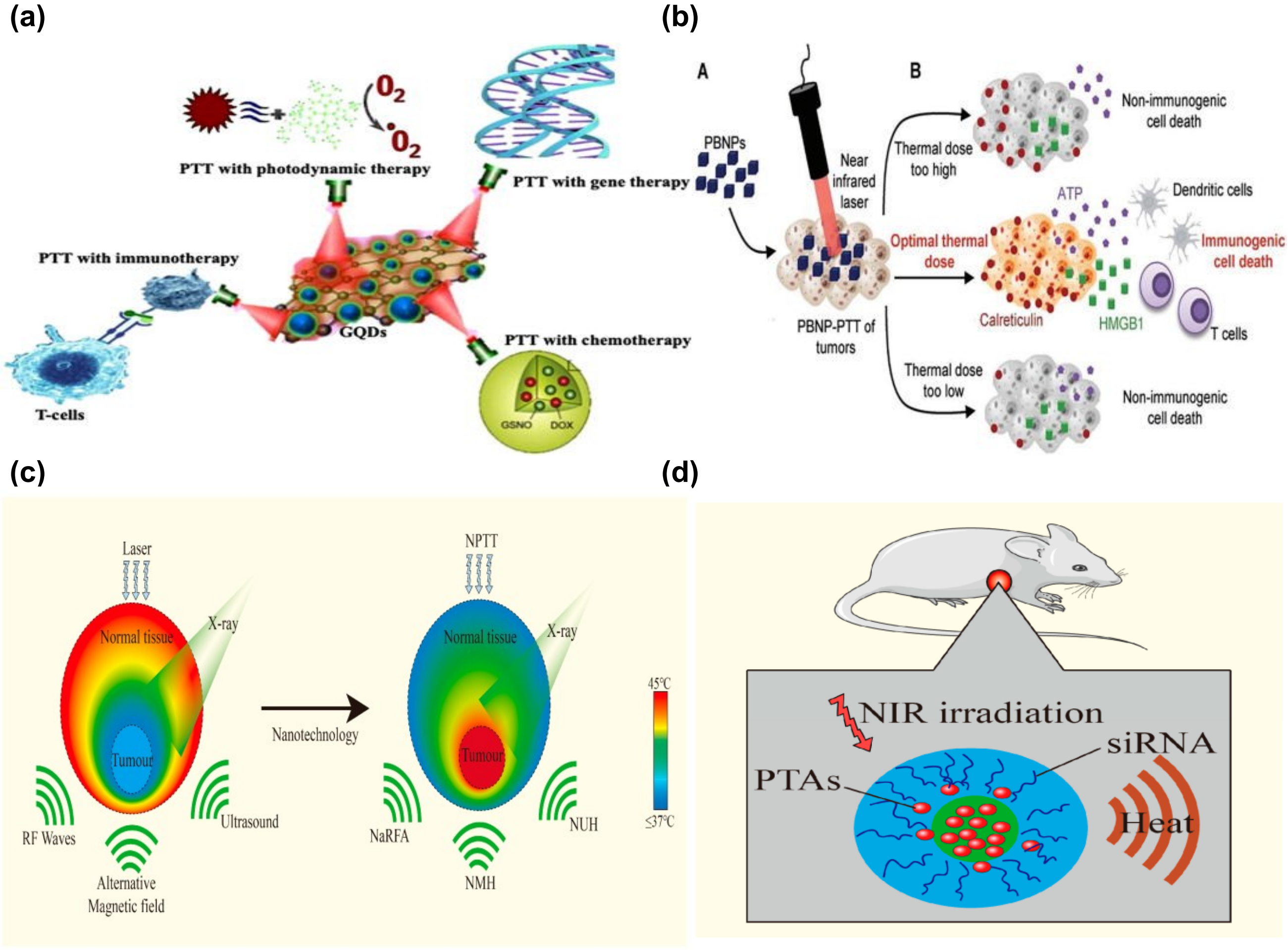

**Figure 9.** Nanophotonics approaches in medicine for photothermal therapy. (a) PTT in combination with different therapeutics.[27] (b) Demonstrates how PTT causes immunogenic cell death.[27] (c) Combination of PTT with Radiotherapy.[151]1 (d) Combination of PTT with Gene Therapy.[151]

### 4.1.2. Two-Dimensional (2D) Materials and Transition Metal Dichalcogenides (TMDs)

Molybdenum disulphide (MoS2) has photothermal absorption (PTA) character. It is a CT imaging agent, has low toxicity, fast biodegradable. Also, it is an excellent NIR absorber with high optical stability and PCE.[152] The MoS2-Ru nanocomposite exhibits synergistic CDT/PTT, a 60.9% apoptotic ratio in TNBC cells, a PCE of 41%, and superior catalytic activity (500 ng/mL compared to 20μg/mL for MoS2 alone).[153] Black Phosphorus (BP) has unique tissue-responsive properties and is biocompatible and biodegradable. HA@BCN is a combination of HA hydrogel with BCN for cutaneous melanoma, which shows>80% cancer cell mortality and tumor inhibition. Also, BP-PTX increases the permeability of cells to chemotherapy. BP nanosheets + R837 is a combination that improves immune activity in PTT-driven cancer immunotherapy. Titanium Carbide (Ti3C2 MXene) has many uses for bone regeneration and antibacterial activity. Ti3C2 + ferrous ions is a combination that improves synergistic PTT-chemodynamic treatment. Also, Ti3C2 nanosheets have concentration-dependent antibacterial activity. Niobium Carbide (Nb2C) is used for penetrating tissue depth for NIR-II PTT. In contrast to the 808-nm laser, the Nb2C-PVP demonstrated deep tissue penetration using a 1063-nm laser. Sharp Nb2C inhibits bacterial biofilm formation and increases heat sensitivity. Also, Iron Phosphoselenide (FePSe3) can be used in immune checkpoint therapy. An example of FePSe3 + APP shows an increase in antibody-labeled T-cell-related immunoreaction after PTT, which is mediated by blocking PD-1.[154]

### 4.1.3. Organic and Other Inorganic Nanomaterials

Beyond noble metals, carbon-based materials, and TMDs, a wide range of additional organic and inorganic nanomaterials are being investigated for PTT, each with its own set of benefits and adding to the diversity of this therapeutic strategy. Organic nanomaterials have higher biocompatibility and biodegradability than many inorganic equivalents, which is critical for reducing long-term toxicity and enabling removal from the body. Polydopamine (PDA) is a commonly utilized component or coating for nanoparticles. It is made by oxidizing dopamine in situ. It has good light absorption and high PCE, making it a popular organic material for PTT.[155] Indocyanine Green (ICG), an FDA-approved photosensitizer used for clinical diagnostics, can cause hyperthermia and reactive oxygen species (ROS) when exposed to NIR lasers. Because of its low stability and fast removal in its original form, ICG is commonly encapsulated into different nanocarriers to increase its bioavailability and therapeutic potential. Examples are mPEG-luteolin-BTZ@ICG for colorectal cancer and ICG-lactosomes for breast cancer. Cypate is a bis-carboxyl-containing indocyanine green derivative that emits fluorescence and produces heat when exposed to NIR light. Cyp-PMMA-Fe@MSCs have been investigated for the diagnosis of lung cancer.

IR-780 is a hydrophobic heptamethine dye that emits more fluorescence than ICG, but is less soluble and quickly removed. Encapsulation in amphiphilic micelle nanoparticles improves stability. Conducting polymers, such as Polyaniline (PANI) and Polypyrrole (PPy), are commonly employed for their high optical absorbance, low cost, and biocompatibility. PANI was one of the first documented polymer-based photothermal anticancer drugs, whereas PPy has a high PCE and improved biocompatibility. Melanin-like polymers are found in plants, animals, and people, have antioxidant, radio-resistant, and anti-neoplastic characteristics, as well as strong NIR radiation absorption. Melanin PEGylated nanoliposomes have shown promise for therapeutic use in skin cancer. Naphthalocyanines and phthalocyanines are chemical dyes designed for strong NIR absorption, reducing light scattering and absorption by biological tissues. Phthalocyanine, for instance, is a well-known organic PTA with a well-defined chemical structure that can be synthesized repeatedly.[149]

On the other hand, inorganic materials such as Semiconductor Materials exhibits excellent NIR absorption and PCE. It includes 2D materials like copper-, bismuth-, and tungsten-chalcogenides (e.g., CuS, Bi2S3), which generally. The characteristics of long-term biocompatibility and degradation patterns are still being investigated. Despite Iron (Fe)- and Manganese (Mn)-based oxides having lower intrinsic PCE, they are widely studied. They have multipurpose potential, like application as MRI contrast agents or magnetic hyperthermia effectors. Iron oxide nanoparticles (IONPs) are degraded by lysosomes into Fe ions, and through the metabolic pathway, they are eliminated.[156]

### 4.1.4. PTT in Combination with Conventional Therapies (Chemotherapy, Radiotherapy, Gene Therapy)

PTT + Chemotherapy improves drug delivery, triggers drug release, overcomes drug resistance, and increases cell membrane permeability. For example, PEG@Pt/DOX and PTX-GO-PEG-OSA improve cytotoxic effects on drug-resistant breast cancer and gastric cancer cells. ZrC nanosheets increase tumor inhibition through prodrug activation and release of chemotherapeutic medicines upon laser irradiation. PTT + Radiotherapy acts through inhibiting DNA repair, cell cycle synchronization, increasing oxidative stress, overcoming radioresistance, and complementary cell killing. For example, Mild Hyperthermia (41-43°C) improves radiosensitivity with a thermal enhancement ratio of (1.5-5.7).[151] In addition, W18O49 nanospheres have excellent radiation sensitization and photothermal performance, suppressing tumor proliferation and metastasis. PTT + Gene Therapy enhances tumor cell apoptosis and improves the delivery of genetic material. In addition, MSC membrane-camouflaged PDA cores demonstrate synergistic chemo-photothermal therapy and gene therapy by delivering siRNA.

### 4.1.5. PTT Synergies with Immunotherapy

This combination enhances mechanisms like immunogenic cell death (ICD), tumor microenvironment (TME) modulation, enhanced immune cell infiltration, and synergy with immune checkpoint blockade (anti-CTLA-4, anti-PD-L1/PD-1). For example, metal-based nanomaterials (Au, Ag, Pt, Pd, TMDCs), when combined with anti-CTLA-4 and anti-PD-L1/PD-1 therapy they enhance anticancer ability.[157] In addition, Corn-like Au/Ag nanorods + anti-CTLA-4 induce an immune memory and prevent tumor recurrence.[158] Also, when nanodrugs are directly or indirectly targeted to the tumour site and release tumor-associated antigens and cell fragments, PTT can cause immunogenic cell death, which activates systemic immunity and eradicates any remaining or metastatic malignancy.[27] Finally, FePSe3 nanosheets + anti-PD-1 prolong survival through promoting dendritic cell maturation and T cell activation.

### 4.1.6. PTT Integration with Other Light-Activated and Emerging Therapies (PDT, CDT, SDT)

PTT + Photodynamic Therapy (PDT) have synergistic mechanisms by overcoming hypoxia, synergistic ROS production, and complementary cell death mechanisms. For example, MWCNT-mTHPC is a combination of PDT and PTT, which induces apoptosis through oxidative stress-mediated and mitochondrial damage in ovarian cancer. In addition, Graphene (GFN) enhances ROS formation for both PDT and PTT. Finally, Covalent organic framework nanoparticles are an example of PTT temperature increase significantly enhanced PDT sensitivity.

PTT + Chemodynamic Therapy (CDT) synergy is achieved through synergistic ROS production via Fenton-like reactions, specifically targeting the TME. In addition, MoS2-Ru nanocomposite, which enhances catalytic activity and photothermal conversion, leads to synergistic CDT/PTT outcomes in TNBC cells. Finally, Ti3C2 nanosheets and ferrous ions provided PTT-chemodynamic synergistic treatment by creating ROS.

PTT + Sonodynamic Therapy (SDT) combination involves combined activation by light and ultrasound, which enhances ROS generation for deep tumors. In addition, CuS-Pt complex improves photothermal performance and catalyzes O2 generation, which improves synergistic tumor killing.

## 4.2. Applications for AI in Nanophotonic Healthcare Devices

AI integration with nanophotonic healthcare devices has major advances in diagnosis and treatments using advanced light engineering at the nanoscale with biomolecules.

### 4.2.1. AI-Enhanced Nanophotonic Diagnostics

Nanophotonic biosensors have the ability for early diagnosis of complex diseases like cancer.[159] They have the ability and sensitivity to detect biomolecules in the absence of labels.[7] AI can provide more signal processing of the optical signals to increase biomarker true sensitivity, specificity, and accuracy.[160] It can identify small shifts in spectra or changes in intensity within one of the SPR sensors. This can yield detection of small amounts of target analytes.[161] In addition, AI can provide many datasets that can help researchers in many disease scenarios.[162] In the COVID-19 pandemic, the integration of AI nanophotonic biosensors helps in fast evaluation of the presence of infection and the early detection of the virus.[163]

AI integration in nanophotonics. Imaging modalities like Optical Coherence Tomography (OCT) provide high-resolution imaging of biological structures.[164] Deep learning models will enhance imaging technologies by improving resolution and reducing image noise, to improve accurate diagnosis.[165] In addition, AI can transform from one imaging modality to another, which helps to extract relevant information from one imaging modality using data from another.[166]

#### 4.2.2. AI-Driven Nanophotonic Therapeutics

Nanophotonics may be used in drug delivery. AI can increase the targeting capabilities of nanoparticles and improve patient treatment outcomes.[167] AI can increase the delivery of drugs by predicting the shape of nanoparticles for maximum distribution within tumor tissues.[168] Nanophotonics can also play a role in therapeutic platforms.[169] AI can improve the nanophotonic agents used in PTT and PDT to either capture more light at particular therapeutic wavelengths, improve heat generation, and increase ROS production, which improves the efficacy of these light therapies.[170]

Nanocarriers are a system for the delivery of drugs or therapeutic genes and small interfering RNA molecules.[171] Gene delivery nanocarriers depend primarily on AI technology to optimize biological stability, promote cellular entry, and enhance target cell recognition. AI improves the outcomes of gene therapy by its analytical capabilities.[172]

## 4.3. The Role of Nanophotonics in Detecting MicroRNA Cancer Markers

Surface plasmon resonance (SPR) and localized surface plasmon resonance (LSPR) sensors is an optical sensing technology that detects small changes in the local refractive index to observe binding of a molecule to a metal surface.[173] Exciting polarized surface plasmons (SPPs) generates the evanescent wave to detect binding events occurring at the surface.[174] Localized Surface Plasmon Resonance (LSPR) is a type of plasmonic sensing that is different from SPR sensing, which does not require prism coupling, and allows for simpler and smaller optical systems. The rainbow of LSPR could detect miRNAs in very low concentrations when combined with methods to amplify the signal, typically at molar amounts.[175, 176]

Photonic crystal (PCs) are periodical nanostructures which manufactured to control the transmission of light, and form optical bandgaps.[177] Using photonic crystal surfaces increases excitation of fluorescently tagged biomolecules and efficiently channels photons to detectors.[178] The combination of photonic crystal arrays and molecular amplification techniques such as CRISPR/Cas12a and HCR can greatly improve the sensitivity of detection for low-abundance miRNAs. More complex designs, such as encoding oriented conformal resonance (GMR) sensors, make the overall system design easier and allow for simultaneous measurements of many analytes, drawing from.[179] Photonic crystal sensors offer opportunities to combine both optical and molecular signal amplification, making them truly remarkable.

Quantum dots (QDs) are semiconductor nanocrystals with excellent optoelectronic properties. It can make excellent fluorescent probes for miRNA detection, and their use is based on the way of specific miRNA binding events.[180] The powerful technique for monitoring miRNA in QDs is called FRET.[181]

## 4.4. Nanophotonics-Enhanced Chemotherapy

Plasmonic Nanoparticles (e.g., Gold, Silver) are Localized Surface Plasmon Resonance (LSPR), and are effective in photothermal conversion, have good photostability, and reduced cytotoxicity. In chemotherapy, they improve the drug delivery by using Light-to-Heat Conversion (PTT). They are also used in tumor ablation, drug injection, and bioimaging contrast improvement.[182] QDs are small in size, have adjustable photoluminescence, narrow emission, high quantum yield, and photobleaching resistance.[183] In chemotherapy, they are used in drug delivery by ROS generation (PDT), bioimaging, photosensitizers, and real-time drug route monitoring.[28] Carbon-based Nanomaterials (e.g., Graphene, Carbon Dots, Nanodiamonds) have large surface area, mechanical strength, electrical/thermal conductivity, tunable surface chemistry, and some intrinsic photosensitizers. In chemotherapy, they are used in drug delivery by using Light-to-Heat Conversion (PTT), ROS Generation (PDT), or Drug Encapsulation.[184] Liposomes are biocompatible, biodegradable, drug encapsulation, sustained release, and temperature-sensitive. They are used in the delivery of chemotherapeutic medications (i.e., Doxil®), and light-triggered drug release (if temperature sensitive). Polymeric Nanoparticles are synthetic polymers, used in medication safety, controlled drug release, and targeted drug delivery (EPR/Active). For example, Paclitaxel/doxorubicin encapsulation for active targeting.[185, 186]

## 4.5. Imaging Modalities and Image-Guided Surgery (IGS) Enhanced by Nanophotonics

For most solid tumors, surgical excision is still the most common and successful therapeutic option.[187] However, reducing the chance of tumor recurrence and enhancing long-term survival depend on obtaining total tumor excision with negative surgical margins, or no malignant tissue left at the corners of the removed specimen. The sensitivity and specificity required to detect microscopic residual disease or precisely define tumor borders in real-time are sometimes lacking in conventional techniques for evaluating surgical margins, such as intraoperative frozen section analysis (IFSA).

Positive surgical margins (PSMs), which require subsequent treatments or reoperations and expose patients to extra side effects and financial stress, are often the result of this.[188] Image-guided surgery (IGS) seeks to get around these significant restrictions by giving surgeons real-time optical imaging input during the process. IGS is defined here as surgery guided with real-time optical imaging feedback. The use of nanophotonics has greatly upgraded numerous essential imaging modalities, increasing their usefulness in minimally invasive and image-guided surgical operations. These advancements are discussed in this section.

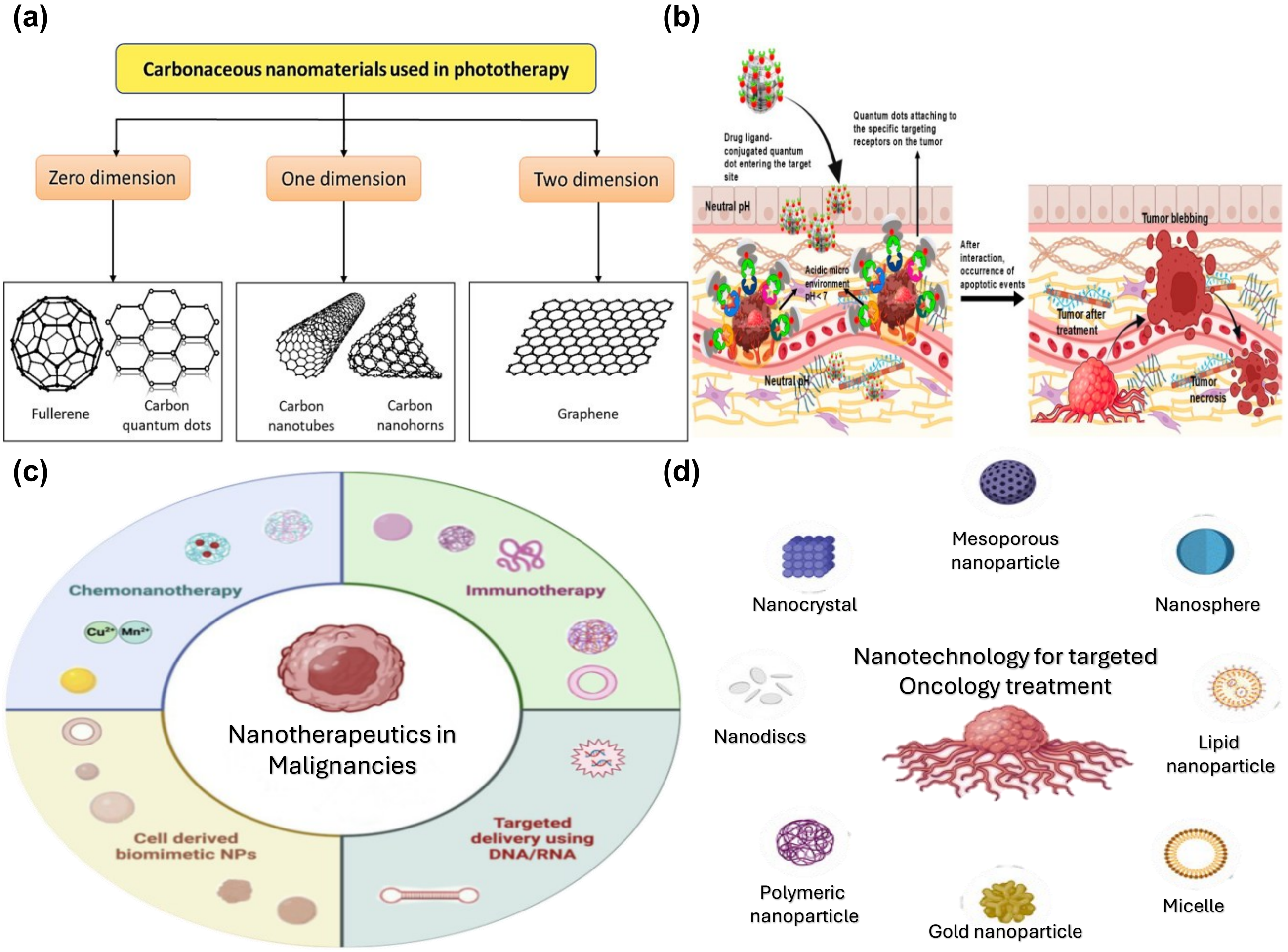

**Figure 10.** Nanophotonics approaches in healthcare for diagnostics and treatment. (a) Dimensions of nanomaterials, especially Carbonaceous.[184] (b) Shows how drug-ligand-conjugated QDs deliver chemotherapeutic medicines with a specific activity.[28] (c) Nanotherapeutics Strategies in Malignancies.[185] (d) Nanotechnology for targeted Oncology treatment.[185]

### 4.5.1. Fluorescence-Guided Surgery (FGS)

(FGS) is a prominent image-guided surgery method that provides surgeons with real-time visual feedback during operations. Although fluorescent image-guided surgery (FIGS) has several advantages over conventional imaging techniques, it has limitations due to shallow tissue penetration and autofluorescence from clinically authorized fluorophores. To solve this issue, researchers are transitioning from visible to longer wavelengths. The NIR spectrum, including NIR-I (700-900 nm) and NIR-II (950-1700 nm), improves signal-to-noise ratios and allows for deeper tissue penetration. Recent breakthroughs in nanophotonics have focused on nanoformulations of NIR-II fluorophores in order to improve in vivo stability and tumor-targeting delivery. Soft NPs containing NIR-II fluorophores are commonly used as contrast agents. PDFT1032 polymeric NPs, for example, were made up of PDFT, a new NIR-II-emitting polymer encased in a DSPE-mPEG shell. PDFT1032 produced high TBRs in osteosarcoma for up to 3 days following an IV injection.[188]

### 4.5.2. Photoacoustic Imaging (PAI)

PAI can be used in direct medication administration, treatment planning, and surgery. It can be used to assess intraoperative dense decalcified and non-decalcified bone samples. This technique might help identify intraoperative tumour margins and detect bone tissue disorders. In addition, Multispectral optoacoustic tomography (MSOT), a kind of PAI, has shown promise for SLN mapping when combined with indocyanine green (ICG). This method helps to rule out metastases. In patients with Crohn's disease, to differentiate between intestinal remission and active disease can use MSOT technique which determines haemoglobin levels and substitutes for Inflammation. Nanoparticles can be used as contrast agents in image-guided surgery.

### 4.5.3. Surface-Enhanced Raman Spectroscopy (SERS)

Recently, SERS has been used in image-guided cancer surgery. Stimulated Raman scattering microscopy has made it possible to image and detect tumours that infiltrate the brain. It has made it easier to obtain diagnostic histology data during brain tumour surgery. It has also shown promise at the preclinical stage for differentiating tissues during tumour surgery.[189]

### 4.5.4. Optical Coherence Tomography (OCT)

OCT has high-resolution imaging and is non-invasive.[190] It provides detailed anatomical structure information. An important advantage is that it provides label-free molecular contrast. OCT has been widely used in different fields like cardiovascular disease, ophthalmology, and skin. It can identify malignancies in a variety of tissues. For example, it helps identify brain tumours. In addition, tissue differentiation and margin assessment. For another example, Optical Coherence Elastography (OCE) has made it possible to identify any remaining cancer in the surgical cavity after breast-conserving surgery (BCS) in vivo.

### 4.5.5. Upconversion Nanoparticles (UCNPs)

The goal of UCNP development is to improve bioimaging. They are used as a technique in image-guided treatments and diagnostics. For example, use in cancer detection, like lymph node metastasis.

Detection of lymph node metastases is a critical component of cancer staging and therapy planning. They also integrate with Theranostic Platforms, and this platform seeks to achieve simultaneous fluorescence imaging and chemo-photodynamic combination therapy.[190]

#### 4.5.6. Quantum Dots (QDs)

QDs are being used for fluorescence-guided cancer surgery and diagnosis. They are also used in vivo in molecular and cellular imaging.[64] In vivo studies have also examined their potential for lymph node imaging. Some varieties, such as PbS quantum dots, provide fluorescence imaging for breast tumours.[187]

# 5. Nanophotonics for Artificial Intelligence and Optical Computing

The exponential growth of data-centric applications and the escalating computational demands of artificial intelligence (AI), particularly deep learning, have exposed fundamental limitations in conventional electronic computing architectures rooted in the von Neumann paradigm. The physical separation of memory and processing units creates an intrinsic bottleneck in data transfer, consuming excessive energy and constraining processing speeds, especially for matrix multiplications and convolutions inherent in neural network operations. Nanophotonics, which explores light-matter interactions at sub-wavelength scales, offers a revolutionary pathway to transcend these limitations. By harnessing photons as information carriers, nanophotonic systems promise ultra-high bandwidth, massively parallel processing capabilities, minimal heat dissipation, and the potential for direct manipulation of optical information (e.g., images, sensor data) without inefficient electro-optical conversions.[191, 192] This section comprehensively examines the cutting-edge advancements in nanophotonics for AI and computing, delving into the underlying principles, diverse architectures, novel materials, and key devices driving progress in optical neural networks (ONNs), neuromorphic photonic computing, and emerging quantum photonic approaches. A critical analysis of the persistent challenges and limitations confronting the field is also presented, alongside insights into ongoing research strategies aimed at realizing the transformative potential of light-based intelligent computing systems.

## 5.1 Optical Neural Networks (ONNs)

Optical Neural Networks (ONNs) represent a direct physical implementation of artificial neural network architectures using photonic components, capitalizing on the inherent parallelism, speed of light propagation, and energy efficiency of optical systems to perform core computational tasks in AI, such as classification, regression, and feature extraction. Nanophotonics provides the essential toolkit for miniaturizing and integrating the fundamental building blocks—artificial neurons and synapses—and efficiently interconnecting them on-chip or in free space.

### 5.1.1 Core Architectures and Platforms

ONNs are primarily categorized based on their physical implementation platform. Free-Space Optical Neural Networks (FSONNs) utilize bulk optics or planar optical elements arranged in three-dimensional space to manipulate light waves representing data. Central to FSONNs are diffractive optical elements (DOEs), particularly multi-layer diffractive structures known as Diffractive Deep Neural Networks ($D^2$NNs) and metasurfaces, where light propagation between layers performs the linear transformations (matrix multiplications) fundamental to neural networks via diffraction and interference. Montes McNeil *et al.* provide a detailed taxonomy of FSONNs, highlighting implementations based on 3D printed layers, dielectric or plasmonic metasurfaces, and spatial light modulators (SLMs).[193] FSONNs excel in massive parallelism and direct handling of large input datasets like full images but often grapple with challenges in reconfigurability, sensitivity to alignment, and large physical footprints. Significant progress was developed through pluggable multitask diffractive neural networks employing cascaded metasurfaces.[194-196] By fixing one metasurface and switching pluggable metasurface modules, their system reconfigured to perform distinct recognition tasks. For example, handwritten digits versus fashion products at near-infrared wavelengths, which exemplifying a pathway towards versatile, high-speed, low-power multifunctional AI systems. Similarly, Tang *et al.* introduced an "Optical Neural Engine" (ONE) architecture that synergistically combines diffractive networks for Fourier space processing with optical crossbar structures for real space processing, enabling efficient and reconfigurable solutions to complex scientific partial differential equations (PDEs).[197]

In contrast, Integrated Photonic Neural Networks (IPNNs) confine and guide light within nanoscale waveguides fabricated on semiconductor substrates like silicon or silicon nitride. Key components include Mach-Zehnder Interferometers (MZIs) and microring resonators (MRRs). MZI meshes can be programmed to implement arbitrary unitary matrices, performing matrix multiplications, while MRRs modulate light intensity through resonance shifts, acting as tunable weights or activation functions. Dong *et al.* extensively chronicle the evolution from free-space to on-chip platforms, emphasizing the roles of MZIs, MRRs, and metasurface-based diffractive networks within photonic circuits.[198] IPNNs offer advantages in compactness, stability, potential for co-integration with electronics, and high operational speeds. However, scalability remains a critical hurdle due to waveguide crossing losses, signal attenuation, and the relatively large footprint of tunable elements like MZIs. Gu *et al.* addressed this with "Squeeze Light," a scalable ONN architecture utilizing multi-operand ring resonators (MORRs) to execute vector dot-products within a single device, significantly enhancing compactness and efficiency compared to conventional MZI-based designs.[199] Qu *et al.* showcased an inverse-designed integrated nanophotonic ONN based on optical scattering units, achieving high-precision stochastic matrix multiplication and image classification (MNIST) within an ultra-compact footprint (4x4 μm$^2$ per unit).[200] Further blurring the lines between guided-wave and free-space approaches, Wang *et al.* explored integrating metasurfaces directly onto silicon photonic platforms, creating on-chip analogues of lenses and spatial light modulators for enhanced signal processing and computing functionalities.[201]

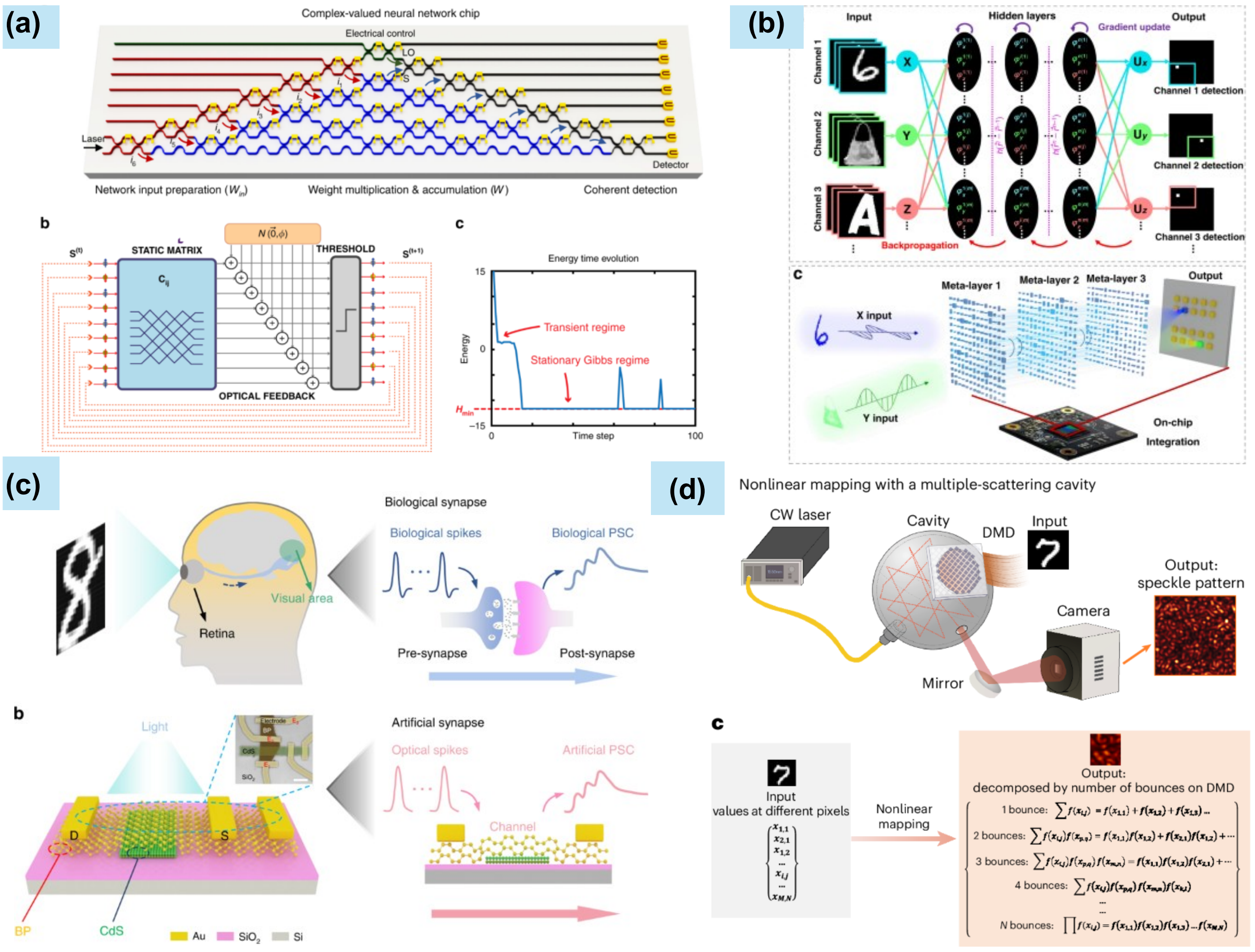

**Figure 11.** Comparative Architectures for AI Computing. (a) Integrated Photonic Neural Network (IPNN) core performing in-memory matrix multiplication using a Mach-Zehnder Interferometer (MZI) mesh or bank of microring resonators (MRRs). Reprinted with permission from Springer Nature.[202] (b) Free-Space Optical Neural Network (FSONN) employing cascaded diffractive layers (e.g., metasurfaces) for massively parallel optical computation. Reprinted with permission from Springer Nature.[203] (c) Neuromorphic Photonic System featuring spiking signals generated by nano-optoelectronic neurons (e.g., RTD-based) communicated via optical interconnects and processed by memristive synaptic crossbars. Reprinted with permission from Springer Nature.[204] (d) Nonlinear optical neural networks featuring optical cavity with multiple internal reflection. Reprinted with permission from Springer Nature.[205]

### 5.1.2 Enabling Multifunctionality and Efficiency with Metasurfaces

Metasurfaces, planar arrays of subwavelength nanostructures (meta-atoms) that impart designed local phase, amplitude, and/or polarization shifts on incident light, have become a cornerstone technology for ONNs. Their subwavelength thickness enables ultra-compact devices and provides unprecedented control over light wavefronts. Within Free-Space $D^2$NNs, metasurfaces are ideal for implementing diffractive layers. By spatially varying the meta-atom design, the precise phase profile required for specific diffraction patterns, representing desired linear transformations, can be encoded. Cheng *et al.* demonstrated such metasurface-based $D^2$NNs for image classification tasks (handwritten digits and animals) using inverse-designed silicon nanodisks, achieving high optical transmittance and classification accuracy (>90% for digits).[206] Luo *et al.* advanced the field with a multi-task optoelectronic hybrid neural network employing a nonlinear metasurface featuring U-shaped resonant units. This design generated second harmonic light, enabling phase multiplexing at both fundamental and second harmonic frequencies, which allowed simultaneous handwritten digit classification (95.53% accuracy) and image reconstruction on the MNIST dataset within a unified architecture.[207]

Moving beyond static implementations, programmable and reconfigurable metasurfaces are crucial for adaptive learning and multifunctional systems. Wang *et al.* demonstrated a multichannel meta-imager utilizing an electrically tunable liquid crystal (LC)-integrated metasurface. By exploiting polarization and angle multiplexing dynamically controlled by applied voltages, the system exponentially increased the number of convolution kernels, performing both positive and negative convolutions simultaneously and achieving high accuracy in image classification (98.5% for digits, 90.9% for fashion images).[208] Ma *et al.* conceptualized "Information Metasurfaces" and "Intelligent Metasurfaces," merging the digital coding/programmable metasurface paradigm with AI for self-adaptive devices, intelligent imagers, and programmable optical AI machines.[209] The integration of metasurfaces with planar photonics is rapidly progressing as Wang *et al.* discussed their deployment on integrated photonic platforms as mode converters and novel light manipulation elements, offering functionalities beyond conventional waveguide devices for computation, imaging, and beam steering.[201]

### 5.1.3 Implementing Linear and Nonlinear Operations

The computational core of neural networks involves linear transformations (matrix multiplications) followed by nonlinear activation functions, both of which must be efficiently realized in ONNs. Linear operations are a natural strength for photonics. Both FSONNs (leveraging diffraction) and IPNNs (using MZI meshes, MRR banks, or scattering units) inherently excel at parallel matrix-vector and matrix-matrix multiplication due to the wave nature of light. Sludds *et al.* proposed a scalable coherent

ONN architecture based on balanced homodyne detection, targeting scalability to millions of neurons with potential sub-fJ/MAC energy consumption, ultimately bounded by photodetector shot noise.[210] Hattori *et al.* implemented optical vector-matrix multiplication (VMM) circuits using wavelength division multiplexing (WDM) for ultra-wideband operation.[211]

Introducing efficient optical nonlinearity, however, remains a significant challenge, crucial for enabling the complex decision boundaries in deep learning. Current strategies primarily involve optoelectronic or all-optical approaches. The optoelectronic method converts optical signals to electrical currents via photodetectors, applies electronic nonlinearity (e.g., using transistors), and then modulates light back using optical modulators. This hybrid approach offers flexibility but incurs latency and energy penalties from optical-to-electrical and electrical-to-optical (O/E/O) conversions. Pursuing all-optical nonlinearity seeks to leverage intrinsic material nonlinearities (e.g., $\chi^{(3)}$ effects like the Kerr nonlinearity in silicon, saturable absorption) or engineered structures (e.g., resonators enhancing nonlinear effects, phase-change materials). Aggarwal *et al.* demonstrated ultrafast switching in antimony (Sb) thin films, suggesting potential for nonlinear elements.[212] Chen *et al.* showcased ultrafast (2 ns) non-volatile photonic memory using Sc-doped $Sb_2Te_3$ (SST) phase-change material, adaptable for nonlinear responses.[213] Several groups consistently identify achieving efficient, low-power, and fast all-optical nonlinearity as a critical frontier for advancing ONNs, potentially involving nanolasers or other active elements as nonlinear neuron sources.[198, 214-216]

## 5.2 Neuromorphic Photonic Computing

Neuromorphic computing aims to emulate the structure (neurons, synapses) and event-driven, low-power information processing principles of biological neural systems. Neuromorphic photonic computing harnesses light to implement these bio-inspired paradigms, offering compelling advantages in speed, bandwidth, and energy efficiency, particularly for spiking neural networks (SNNs) and related models.

### 5.2.1 Photonic Memristive Synapses and Neurons

Memristors, resistive switching devices possessing inherent memory, are ideal candidates for artificial synapses due to their ability to store synaptic weights as conductance states and implement synaptic plasticity rules. Research explores diverse nanoscale material systems for neuromorphic photonics. Oxide-based materials (e.g., $HfO_2$, $TaO_x$) are prevalent in CMOS-compatible resistive RAM (RRAM). Mikhaylov *et al.* summarized CMOS-integrated memristive arrays (RRAM) for neuromorphic computing, emphasizing their use in crossbar arrays for efficient vector-matrix multiplication and the potential for orders-of-magnitude gains in performance and energy efficiency over traditional hardware.[217] Gayakvad *et al.* focused on spinel ferrites (e.g., $CoFe_2O_4$, $NiFe_2O_4$) synthesized via spin coating for RRAM, discussing their resistive switching properties, endurance, and suitability for neuromorphic computing and hardware security applications.[218] Phase-Change Materials (PCMs) like $Ge_2Sb_2Te_5$ (GST) and Sc-doped $Sb_2Te_3$ (SST) offer non-volatile, multi-level switching. Chen *et al.* demonstrated ultrafast neuromorphic photonic memory using SST with 2 ns write/erase speeds, showcasing multilevel capability, stability, and application as synapses in an ANN for image classification, alongside potential for reflective nanodisplays.[213] Two-dimensional materials (graphene, transition metal dichalcogenides - TMDCs) and nanostructures offer unique electronic and optical properties. Sun et al. reviewed synaptic devices (memristors, transistors) based on various nanomaterials (quantum dots, nanowires, 2D materials, oxides, ferroelectrics, organics) for neuromorphic computing.[219] Panes-Ruiz et al. specifically reviewed carbon nanomaterial-based memristive devices (fullerenes, carbon nanotubes, graphene) for neuromorphic applications.[220] Hassanzadeh discussed the broader potential of 2D nanoelectronic materials in bio-inspired computing, including neuromorphic functions.[221] Perovskites represent another promising class; Ma et al. highlighted their use in optoelectronic synapses in addition to demonstrating $CsPbBr_3$ nanoplate-based synapses emulating essential functions like short-term/long-term plasticity and learning-experience behaviour, even featuring a unique memory backtracking capability.[26, 222]

The operation of photonic memristive devices involves setting the conductance state (synaptic weight) using electrical pulses, optical pulses, or a combination of stimuli (heterostimuli). Kim *et al.* exemplified heterostimuli chemo-modulation in ZnO/polyvinylpyrrolidone nanocomposite memristors, combining electrical switching with photostimuli-modulated redox chemistry for associative learning. This device exhibited rapid learning (1 ms), good retention, and, when integrated into an ANN crossbar, enabled highly data-efficient machine learning with exceptional power efficiency.[223] Sun *et al.* positioned memristor-based artificial chips as core components for future brain-inspired AI systems due to their innate in-memory computing capability, tracing their evolution from synapses to neural networks and brain-like chips.[224] Luan *et al.* emphasized the reciprocal development between ANNs and nanophotonics, where neural network algorithms like inverse design facilitate the creation of novel nanophotonic neuromorphic devices.[214]

### 5.2.2 Spiking Neural Networks and Event-Driven Processing

Spiking Neural Networks (SNNs) process information based on the precise timing of discrete spikes (events), closely mimicking biological neural communication and offering high energy efficiency for sparse data. Implementing photonic SNNs necessitates artificial spiking neurons and efficient optical spike communication channels. Emulating spiking neurons leverages various nanophotonic approaches. Nano-opto-electronic devices, particularly Resonant Tunneling Diodes (RTDs) exhibiting folded negative differential resistance (NDR), enable intrinsic spiking behaviour. Romeira *et al.* and Jacob *et al.* are pioneering brain-inspired nanophotonic spike-based devices using III-V nanoRTDs as high-speed artificial neurons.[225-227] Integrating these nanoRTDs with nanoscale light-emitting diodes (nanoLEDs) or nanolasers creates spiking emitter nodes, while integration with nanoscale photodetectors (nanoPDs) forms spiking receiver nodes. Semiconductor lasers subjected to optical feedback exhibit complex dynamics, including excitable (neuron-like spiking) and chaotic regimes, useful for reservoir computing and potentially direct spiking emulation. Wu *et al.* reviewed intelligent optical computing based on laser cavities, covering dynamics relevant to SNNs. Other active devices like VCSELs and microlasers with thresholding and relaxation oscillations can also be engineered for spiking.[228]

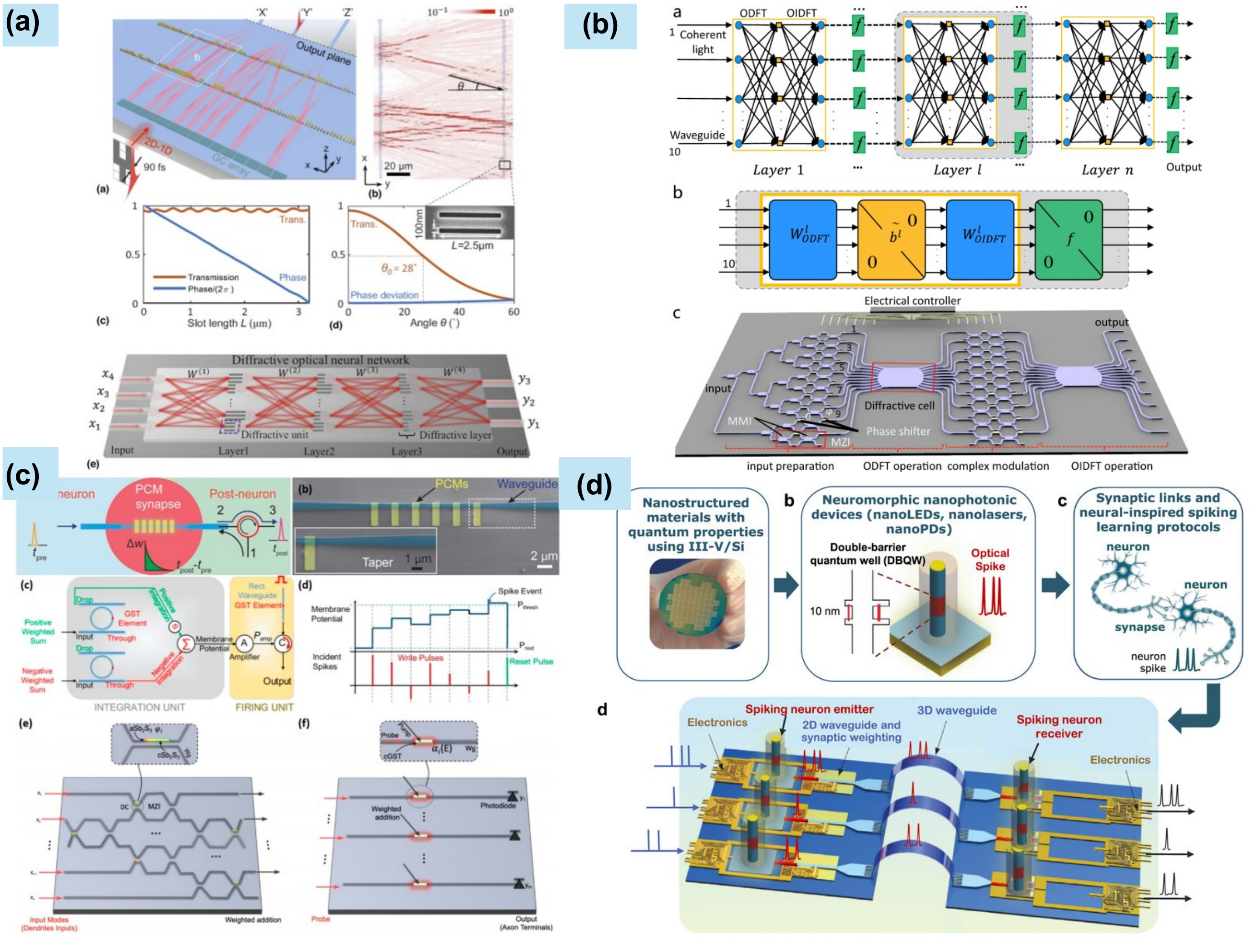

**Figure 12.** Key Nanophotonic Building Blocks for Artificial Intelligence. (A) Metasurface unit cell enabling precise wavefront control for $D^2$NNs or on-chip light manipulation. Reprinted with permission from Springer.[26] (B) Essential integrated photonic components: Mach-Zehnder Interferometer (MZI) for linear algebra, Microring Resonator (MRR) for weighting/filtering, and the challenge of waveguide crossings. Reprinted with permission from Springer.[29] (C) Phase-Change Material (PCM) cell acting as non-volatile photonic memory element or synaptic weight. Reprinted with permission from MDPI.[229] (D) Nano-Opto-Electronic Spiking Neuron concept integrating a Resonant Tunneling Diode (RTD) with a nano-light-emitting diode (nanoLED) or nano-photodetector (nanoPD) for optical spike generation/detection. Reprinted with permission from IOP.[230]

Efficiently routing optical spikes between neurons is critical. Romeira *et al.* discuss employing silicon photonics interconnects, integrated photorefractive interconnects, and 3D polymeric waveguide interconnections. Synchronization of photonic spiking neurons, essential for coordinated computation, is an active research area, often exploiting the nonlinear dynamics and coupling between devices. Li *et al.* reviewed spin-wave excitation and synchronization in Spin Hall Nano-Oscillators (SHNOs), highlighting their potential as nanoscale spiking signal sources for magnonic neuromorphic systems.[231]

### 5.2.3 Photonic Reservoir Computing

Reservoir Computing (RC) is a neuromorphic paradigm utilizing a fixed, randomly connected recurrent network (the "reservoir") with nonlinear nodes to transform input signals into a high-dimensional state space. Only a simple readout layer, typically linear, requires training, simplifying learning for temporal tasks. Photonics is well-suited for implementing the reservoir due to inherent dynamics and parallelism. Common implementations involve delayed feedback systems, where a single nonlinear node (e.g., a semiconductor laser, modulator, or photodetector) with a delayed feedback loop creates virtual nodes in time. Brunner *et al.* pioneered this using a semiconductor laser, with Wu *et al.* reviewing subsequent advances employing various lasers (semiconductor ring lasers, microchip lasers, VCSELs) and photonic integrated circuits.[228] Spatial photonic reservoirs consist of networks of coupled photonic nodes (e.g., microring resonators, photonic crystal cavities) implemented on-chip. Dan *et al.* included RC within broader discussions of intelligent photonic and analog optical computing. RC excels in applications like time-series prediction, speech recognition, and chaotic system modelling, benefiting significantly from the high speed and bandwidth of photonic realizations.[232]

## 5.3 Quantum Photonics for AI Applications

While classical nanophotonics underpins current AI hardware research, quantum photonics leverages quantum mechanical phenomena (superposition, entanglement) and holds long-term promise for specific AI applications demanding exponential speedup or processing inherently quantum data. This field remains nascent regarding direct nanophotonic hardware for mainstream AI. Quantum Neural Networks (QNNs) are theoretical models proposing the use of quantum systems (entangled photons, superconducting circuits) to represent and process information, potentially accelerating certain machine learning algorithms. Implementing QNNs necessitates stable, scalable quantum photonic hardware: integrated sources of entangled photons, low-loss reconfigurable quantum photonic circuits (using components like MZIs and phase shifters), and efficient single-photon detectors. Progress in integrated quantum photonics forms the essential foundation, though specific nanophotonic AI hardware demonstrations are limited. Concepts from quantum computing, like quantum walks or Ising models, can sometimes be mapped onto classical photonic hardware for potential acceleration in optimization and sampling tasks. Wu *et al.* discussed photonic Ising machines using injection-locked laser networks or degenerate cavity lasers to solve combinatorial optimization problems by finding the ground state of the Ising Hamiltonian.[228] Yao and Zheng also noted the use of nanophotonic systems for simulating quantum-inspired models.[215, 216] Quantum photonics is fundamentally required for processing quantum information itself, relevant for future quantum machine learning algorithms operating on quantum data, demanding the same advanced nanophotonic components as QNNs. Zajac *et al.* work on optically controlling adaptive nanoscale domain networks, while classical, illustrates the precision control relevant to quantum systems.[233] Significant challenges persist, including decoherence, high error rates, scalability of quantum hardware, the development of practical quantum algorithms with proven advantage for AI tasks, and the complex integration of quantum photonics with classical control electronics.

## 5.4 Challenges and Limitations in AI and Computing Applications

Despite the compelling promise and substantial progress, the practical realization and widespread deployment of nanophotonic AI and computing systems face significant, multifaceted challenges demanding sustained research efforts. Integration complexity and scalability constitute a primary hurdle. Most practical systems necessitate hybrid photonic-electronic integration, requiring efficient, high-bandwidth, and low-energy interfaces between photonic computing cores (ONNs, neuromorphic arrays) and electronic control logic, memory, and digital processing. Achieving this through monolithic integration (e.g., silicon photonics with CMOS electronics) or advanced heterogeneous integration techniques presents substantial engineering challenges.[198, 226, 227] Scaling IPNNs to millions of neurons intensifies issues like waveguide crossing losses, crosstalk, thermal crosstalk between tuning elements (e.g., heaters on MRRs/MZIs), and the physical footprint of components. Strategies like Gu *et al.* MORRs and Qu *et al.* inverse-designed scattering units aim to mitigate footprint limitations,[199, 200] while Wang *et al.* and Lian *et al.* discuss scaling challenges with metasurface integration and photonic memories, respectively.[201, 234] Exploiting three-dimensional integration, as in some FSONNs or proposed stacked photonic chips, can enhance density but introduces formidable fabrication and alignment complexities.

Achieving strong, fast, low-energy, and scalable all-optical nonlinearity remains a fundamental bottleneck. While optoelectronic solutions provide a workaround, the inherent latency and energy penalty of O/E/O conversion limit ultimate performance. Material innovations including novel semiconductors, 2D materials with giant nonlinearities, epsilon-near-zero materials, optimized PCMs/Antimony and device engineering strategies like resonant enhancement are critical pathways in progress. Programming, calibration, and control present another layer of difficulty. Maintaining precision and stability against thermal drift, fabrication variations, and environmental fluctuations is challenging, particularly for large-scale matrix operations. Techniques like robust training algorithms and on-chip monitoring/feedback are essential, while Sludds *et al.* note fundamental limits like shot noise.[210] Reconfigurability speed is often hampered by slow tuning mechanisms (e.g., thermal heaters on MRRs); faster electro-optic or all-optical tuning is desirable but difficult. The electronic control overhead for managing thousands or millions of photonic elements can itself become a bottleneck in energy, area, and complexity, potentially offsetting the photonic core's advantages.[209]

Training methodologies adapted to photonic hardware are crucial. In-situ training, directly on the physical hardware, is highly desirable but complex due to the difficulty of obtaining gradients through the optical system. Techniques for backpropagation through optical hardware require sophisticated measurement or estimation schemes.[200] Hardware-aware training, run offline but accounting for the specific imperfections, noise, and limitations of the target photonic platform, is vital for achieving robust deployed performance.[235, 236] Achieving true system-level energy efficiency requires minimizing all contributors: optical losses (propagation, coupling, scattering), tuning energy (especially thermal), O/E/O penalties in hybrid systems, and peripheral electronics energy (ADCs, DACs, control logic). Demonstrating end-to-end efficiency surpassing optimized electronic ASICs (e.g., TPUs) remains a key goal.

Maximizing the potential of nanophotonics necessitates algorithm-hardware co-design. Developing AI algorithms specifically tailored to leverage photonic strengths (massive parallelism, analog processing, optical convolutions) while respecting constraints (limited precision, specific nonlinearities, noise) is essential, rather than merely porting existing digital algorithms. Finally, material and fabrication challenges persist. Reliably developing and fabricating materials with required optical properties (low loss, high nonlinearity, fast switching), stability, and CMOS compatibility (e.g., GST, Sb, Sc-Sb-Te, novel oxides, 2D materials) at scale is non-trivial. Fabrication precision and yield for intricate nanophotonic structures (metasurfaces, inverse-designed devices, dense waveguides) directly impact performance, cost, and commercial viability.[237]

# 6 Comparative Analysis

Nanophotonics has shown significant promise in enhancing green energy production, particularly through the development of advanced solar cells. Self-assembled nanophotonic structures have been utilized for light-trapping in thin-film solar cells, which improves the performance and cost ratio of photovoltaics by efficiently manipulating radiated electromagnetic energy.[238] Additionally, nanophotonics enables the optimization of optical, opto-electrical, and thermal responses in materials used for energy-efficient buildings, such as coatings and composites for windows, roofs, and walls.[239] These advancements contribute to sustainability by reducing energy consumption and enhancing the efficiency of renewable energy sources.

Biosensing has revolutionized using nanophotonics by enabling the meticulous detection of subwavelength light, which enhances the precision of bioanalytical measurements. Recent advances in nanophotonic biosensors include phase-driven sensors, resonant dielectric nanostructures, plasmonic nanostructures, surface-enhanced spectroscopies, and evanescent-field sensors. These technologies improve sensor performance and efficacy, addressing challenges such as controlling biological specimens and reducing costs for broader accessibility.[7, 32, 240] The integration of nanophotonics in biosensing is crucial for applications in health and safety, providing high sensitivity and specificity in detecting biological materials.

Medicine and healthcare improved from using nanophotonics by facilitating non-invasive diagnostics and therapeutics through the use of nanophotonic materials like black phosphorous and gold nanoparticles, which offer tunable optical properties for precise targeting and treatment. These materials are used in photodynamic therapy, bioimaging, and multimodal imaging, providing new avenues for effective cancer treatment.[32, 241, 242]

Nanophotonic technologies have significant applications in optical computing, particularly in optical neural networks and neuromorphic computing. As Nanophotonics offer promising methods to realize high-performance computing systems by utilizing the unique properties of light propagation, including multi-band operation, high speed, and low power consumption.[234, 243, 244] Additionally, the development of nanophotonic devices has facilitated the creation of optical logic gates and cryptographic circuits, further advancing the capabilities of optical computing and information processing.[245-247] Overall, the reciprocal development between artificial neural networks and nanophotonics is driving innovations in optical computing, enabling new functionalities and enhancing performance in various applications.

**Table 2.** Summary of Nanophotonics applications

| Area | Application | Reference |
|---|---|---|
| Solar Energy | Perovskite Photovoltaic | 55-58, 60-62, 64, 65, 68, 73 |
| | Concentrated solar power | 24, 80-84 |
| | Thermophotovoltaic | 93, 94, 96-100 |
| Biosensing | Biomolecule detection | 7, 32, 240 |
| | Environmental Monitoring | 110, 112 |
| | Food safety | 128, 129 |
| Medicine | Photothermal therapy | 143-146, 148-150 |
| | Chemotherapy | 182-186 |
| | Image-Guided Surgery | 187-189 |
| Healthcare | Tumour diagnostics | 161-164 |
| | Imaging modality | 187, 190 |
| | mRNA detection | 173, 175, 176, 179-181 |
| Optical computing | Optical Neural Networks | 198, 214-216, 234, 243, 244 |
| | Neuromorphic Computing | 26, 222, 232, 245-247 |

# 7 Conclusion and Outlook

This review has synthesized recent breakthroughs in nanophotonics, underscoring its pivotal role in advancing green energy, precision medicine, biosensing, and optical computing. Nanophotonic engineering enables unprecedented control over light at subwavelength scales, driving innovations such as perovskite solar cells with >30% efficiency, plasmonic biosensors capable of single-molecule detection, and tumor-selective photothermal therapies. In computing, integrated photonic neural networks and metasurface-based platforms offer transformative gains in speed and energy efficiency, potentially overcoming fundamental limitations of electronic systems. Despite these advances, key hurdles impede large-scale commercialization. Material stability remains critical, particularly for perovskite photovoltaics under operational stresses and plasmonic nanostructures susceptible to oxidation. Fabrication complexity and cost hinder scalability; techniques like e-beam lithography for SERS substrates or cleanroom-dependent photonic circuits are economically prohibitive for mass production. System integration, especially coupling nanophotonic components with electronic control units, demands sophisticated heterogenous packaging, increasing design overhead. Additionally, standardization of performance metrics and regulatory frameworks for biomedical applications is underdeveloped. Allowing commercialization of Nanophotonic technology depends on addressing existing challenges through material innovation. By developing robust, eco-friendly nanomaterials (e.g., lead-free perovskites, oxidation-resistant plasmonic alloys) and hybrid systems (e.g., MOF-encapsulated nanoparticles) that balance performance with stability. Then, merging these materials with manufacturing advancements such as scaling low-cost techniques like nanoimprint lithography for metasurfaces, roll-to-roll processing for solar cells, and 3D printing for biosensors to supplant expensive high-precision methods. Concurrently, AI co-design must leverage machine learning for inverse nanostructure design, fabrication optimization, and real-time sensor analytics to enhance performance while curtailing prototyping costs. Multidisciplinary integration is equally vital, converging photonics, electronics, and biotechnology to develop "lab-on-chip" diagnostics and energy-efficient photonic accelerators compatible with existing infrastructure. Finally, establishing standardized testing protocols and incentivizing public-private partnerships through supportive regulatory and economic models will de-risk scaling pilot technologies like nanophotonic thermophotovoltaics and implantable biosensors. As these efforts mature, coordinated research in scalable manufacturing and cross-sector collaboration will transition nanophotonics from laboratory breakthroughs toward enabling sustainable energy grids, accessible diagnostics, and ultra-efficient computing systems.

## Acknowledgement

The authors would like to acknowledge and thank Egypt Scholars organization and specially the Advanced Labs for organizing the team-work to produce this paper.